\documentclass[aps,prd,twocolumn,nofootinbib,superscriptaddress,floatfix]{revtex4-2}
\usepackage{amsmath,amssymb,amsfonts}
\usepackage{subcaption}
\usepackage{graphicx}
\usepackage{hyperref}
\usepackage{xcolor}
\usepackage{bm}
\hypersetup{colorlinks=true,linkcolor=blue,citecolor=blue,urlcolor=blue}
\usepackage{booktabs}
\DeclareUnicodeCharacter{2032}{\ensuremath{^{\prime}}}
\usepackage{fontawesome}
\usepackage{hyperref}
\begin{document}

\title{Theoretical and Experimental Constraints in the $\mu$--$\tau$ Four-Lepton Sector of the SMEFT: implications to neutrino self interactions}
\author{Aadarsh Singh}
\affiliation{Centre for High Energy Physics, Indian Institute of Science, C V Raman Avenue, Bangalore 560 012, India}
\author{Giancarlo D'Ambrosio}
\affiliation{INFN-Sezione di Napoli, Complesso Universitario di Monte S. Angelo, Via Cintia Edificio 6, 80126 Napoli, Italy}
\affiliation{Centre for High Energy Physics, Indian Institute of Science, C V Raman Avenue, Bangalore 560 012, India}
\author{Sudhir K Vempati}
\affiliation{Centre for High Energy Physics, Indian Institute of Science, C V Raman Avenue, Bangalore 560 012, India}
\date{\today}

\begin{abstract}
We study the dimension-six SMEFT four-lepton operators in the $\mu$--$\tau$ 
sector. These operators control both charged-lepton scattering and neutrino 
self-interactions, the latter being weakly constrained by direct laboratory 
probes despite their importance for cosmological tensions. We compare three 
classes of constraints on the Warsaw-basis coefficients $[C_{\ell\ell}]_{2222}$, 
$[C_{\ell\ell}]_{2233}$, and $[C_{\ell\ell}]_{2332}$. We use perturbative 
unitarity from $2\!\to\!2$ partial-wave analysis, spin-summing positivity 
sum rules, and the experimental bounds from NA64$\mu$ and the global 
fit~\cite{Falkowski:2017pss}. We find that the global fit dominates for 
$[C_{\ell\ell}]_{2222}$ and $[C_{\ell\ell}]_{2332}$, while NA64$\mu$ provides 
the leading bound on $[C_{\ell\ell}]_{2233}$, with the unitarity line for this 
direction entering the range of collider energies near $200~\mathrm{GeV}$. 
Renormalization-group running between $1~\mathrm{GeV}$ and $30~\mathrm{TeV}$ 
modifies these coefficients by up to $10\%$. Translating these bounds onto the 
effective four-neutrino coupling $G_\mathrm{eff}$, we find values many orders 
of magnitude smaller than the strongly interacting regime motivated by the 
Hubble tension; this excludes heavy-mediator UV completions of strong $\nu_{\mu}$--$\nu_{\mu}$ and 
$\nu_\mu$--$\nu_\tau$ self-interactions within the validity of the 
dimension-six SMEFT and in the absence of tuned cancellations between 
operators, while leaving the cosmologically motivated light-mediator 
scenarios unconstrained by this analysis. Finally, 
we comment on the bounds these coefficients place on a leptophilic 
$L_\mu - L_\tau$ $Z'$ UV completion. Our SMEFT-based current and projected 
NA64$\mu$ bounds reproduce the dedicated $Z'$ analyses already available in 
the literature. We also present a complementary analysis for the $e$--$\mu$ 
sector using projected JUNO sensitivities.
\end{abstract}

\maketitle

\section{Introduction}
\label{sec:intro}

Neutrino-neutrino scattering is one of the least tested predictions of the Standard Model (SM). The SM cross section for this process is extremely small, set by the Fermi constant $G_F \sim 10^{-11}\;\text{MeV}^{-2}$. Direct laboratory measurements of neutrino self-interactions are essentially nonexistent. Yet this process plays a potentially important role in early-universe cosmology. Strongly self-interacting neutrinos with an effective coupling $G_{\rm eff} \sim 10^{-4}\text{--}10^{-2}\;\text{MeV}^{-2}$, far larger than $G_F$, have been proposed as a possible way to alleviate the Hubble tension between CMB-inferred and local measurements of $H_0$~\cite{Kreisch:2019yzn,Planck:2018vyg,Riess:2021jrx,brinckmann2021self}. Such interactions delay neutrino free-streaming and shift the phase of the acoustic peaks in the CMB power spectrum. They have also been suggested to ease the milder tension in the amplitude of matter fluctuations $\sigma_8$~\cite{Blinov:2019gcj,Escudero:2018mvt}. Several groups have studied these scenarios in detail~\cite{lyu2021self,Berryman:2022hds,das2022selfinteracting,roychoudhury2021updated}. An important finding is that the flavor-universal case, where all neutrino flavors self-interact equally, is strongly constrained by laboratory data. The scenario with self-interactions only among tau neutrinos turns out to be the least constrained and potentially consistent with a solution to the $H_0$ problem~\cite{Blinov:2019gcj,lyu2021self,brinckmann2021self,Berryman:2022hds}. The cosmologically preferred parameter space continues to evolve as Planck-likelihood refinements tighten the free-streaming bounds, with residual allowed windows that depend on degeneracies with $N_{\rm eff}$ and $Y_p$. Closing the laboratory gap in the third-generation lepton sector --- the focus of this paper --- is therefore directly motivated by these cosmological scenarios.

The Standard Model Effective Field Theory (SMEFT)~\cite{Weinberg:1979sa,Buchmuller:1985jz,Grzadkowski:2010es,Brivio:2017vri} provides a natural framework to connect neutrino self-interactions to other observables. At dimension six in the Warsaw basis~\cite{Grzadkowski:2010es,Dedes:2017zog}, the four-lepton operator
\begin{equation}
[\mathcal{O}_{\ell\ell}]_{prst} = (\bar\ell_p\gamma^\mu \ell_r)(\bar\ell_s\gamma_\mu \ell_t)
\label{eq:Oll}
\end{equation}
simultaneously generates both charged-lepton and neutrino scattering amplitudes, up to electroweak symmetry breaking effects and renormalization-group mixing into other four-fermion structures. This is because the left-handed lepton doublet $\ell_p$ is an $SU(2)_L$ multiplet containing both the charged lepton and the neutrino of flavor $p$. So a bound on the Wilson coefficient from charged-lepton processes automatically translates into a constraint on the corresponding neutrino channel~\cite{Falkowski:2017pss,Falkowski:2019hvp}. Global SMEFT fits~\cite{Ellis:2020unq,Ethier:2021bye,Berthier:2015oma,Falkowski:2017pss} extract bounds on these Wilson coefficients from a wide variety of processes~\cite{ParticleDataGroup:2022pth,Crivellin:2020ebi}. This indirect strategy is currently the most powerful way to constrain neutrino self-interactions at the dimension-six level.

However, the experimental reach is not uniform across the three leptonic flavor sectors. The $e$--$\mu$ and $e$--$\tau$ sectors benefit from abundant $e^+e^-$ collider data. LEP measurements of $e^+e^-\to\mu^+\mu^-$ and $e^+e^-\to\tau^+\tau^-$ at and above the $Z$ pole, supplemented by neutrino-trident production~\cite{Altmannshofer:2019zhy,Altmannshofer:2014pba,CCFR:1991lpl} and low-energy precision tests~\cite{LEP:2003aa,Efrati:2015eaa,Falkowski:2017pss,Corbett:2014ora}, already constrain the corresponding Wilson coefficients at the $\mathcal{O}(10^{-2})$~TeV$^{-2}$ level or better. The global SMEFT fit projections of Ref.~\cite{deBlas:2022qbs} show that future $e^+e^-$ facilities such as FCC-ee, CEPC, ILC, and CLIC~\cite{FCC:2018evy,cepc2018cepc,Bambade:2019fyw,CLICdp:2018cto} will tighten these bounds by one to two orders of magnitude, owing to the enormous statistics in $e^+e^-\to f\bar f$ processes at high luminosity. The $\mu$--$\tau$ sector tells a very different story. The four-lepton operators $[C_{\ell\ell}]_{2233}$ and $[C_{\ell\ell}]_{2332}$ do not enter $e^+e^-\to f\bar f$ at tree level. So these future $e^+e^-$ colliders offer essentially no improvement for the $\mu$--$\tau$ coefficients~\cite{deBlas:2022qbs}. The experimental gap between the well-tested $e$--$\mu$ and $e$--$\tau$ sectors and the $\mu$--$\tau$ sector will only widen with the next generation of $e^+e^-$ machines. The only avenues for closing it are NA64$\mu$~\cite{NA64:2024nwj,crivelli2025first} and a future muon collider~\cite{Accettura:2023ked,AlAli:2021let,MuonCollider:2022xlm,Han:2020dwo,Buttazzo:2020uzc}.

A similar hierarchy persists in the neutrino oscillation sector. Reactor experiments such as JUNO~\cite{adam2015juno,an2016juno} are primarily sensitive to $\bar{\nu}_e$ through inverse beta decay, while near detectors like TAO~\cite{abusleme2020tao} provide precise measurements of the unoscillated $\bar{\nu}_e$ spectrum. Long-baseline experiments such as DUNE~\cite{abi2020deep} and T2HK~\cite{abe2018hyper} probe $\nu_\mu \to \nu_e$ appearance and are mainly sensitive to flavor combinations involving electrons. Atmospheric neutrino experiments including IceCube~\cite{aartsen2018measurement} and KM3NeT~\cite{adrian2016letter,km3net2021orca} offer access to $\nu_\mu \leftrightarrow \nu_\tau$ oscillations and provide the leading sensitivity to the $\mu$--$\tau$ sector, but their reach remains weaker than the constraints involving first-generation leptons. This pattern arises from the dominance of electrons in matter effects and the reliance of most detection channels on $\nu_e$ interactions. Even in future projections, these experiments improve sensitivity mainly to electron-involving combinations, with limited reach in the pure $\mu$--$\tau$ sector. This is consistent with the SMEFT-to-NSI analysis of Ref.~\cite{du2022exploring}.

In this context, we focus on the three flavor-conserving $\mu$--$\tau$ operators in the Warsaw basis,
\begin{align}
[\mathcal{O}_{\ell\ell}]_{2222}&=(\bar\ell_2\gamma^\mu\ell_2)(\bar\ell_2\gamma_\mu\ell_2), \\
[\mathcal{O}_{\ell\ell}]_{2233}&=(\bar\ell_2\gamma^\mu\ell_2)(\bar\ell_3\gamma_\mu\ell_3), \\
[\mathcal{O}_{\ell\ell}]_{2332}&=(\bar\ell_2\gamma^\mu\ell_3)(\bar\ell_3\gamma_\mu\ell_2).
\end{align}
These operators are defined in the Warsaw basis; equivalent representations may differ by Fierz transformations and identities specific to the chosen basis~\cite{cao2025unitarity}. Expanding the $SU(2)_L$ doublets, $[\mathcal{O}_{\ell\ell}]_{2222}$ generates the charged-lepton process $\mu^+\mu^-\to\mu^+\mu^-$, the neutrino self-interaction $\nu_\mu\nu_\mu\to\nu_\mu\nu_\mu$, and the mixed channel $\mu\,\nu_\mu\to\mu\,\nu_\mu$. Similarly, $[\mathcal{O}_{\ell\ell}]_{2233}$ and $[\mathcal{O}_{\ell\ell}]_{2332}$ yield the analogous tau and $\mu$--$\tau$ mixed channels together with their neutrino counterparts. The coefficient $[C_{\ell\ell}]_{3333}$ remains essentially unconstrained by direct experiment unless one imposes a $U(3)^5$-like flavor symmetry~\cite{Falkowski:2017pss,Remmen:2020uze}, and we do not discuss it further here. Some of these $\mu$--$\tau$ coefficients have recently been constrained for the first time by the NA64$\mu$ experiment at CERN~\cite{NA64:2024nwj,crivelli2025first,Gninenko:2014pea}, which probes missing-energy signatures from muon beams on a fixed target. These results provide direct bounds on the hatted combination $\hat{C}_{\ell\ell}^{2222}$ and on $[C_{\ell\ell}]_{2233}$ without relying on flavor-symmetry assumptions. The hatted combination is defined in terms of the SM gauge couplings $g_Y$ and $g_L$ and the Wilson coefficients $[C_{\ell\ell}]$ and $[C_{\ell e}]$ as $\hat C_{\ell\ell}^{2222}\equiv[C_{\ell\ell}]_{2222}+\frac{2 g_Y^2}{g_L^2+3g_Y^2}[C_{\ell e}]_{2222}$.

On the theory side, these Wilson coefficients are subject to constraints from perturbative unitarity and analyticity of the $S$-matrix. We follow the approach of Ref.~\cite{bresciani2026unitarity} to apply the partial-wave unitarity bounds on the $\mu$--$\tau$ coefficients~\cite{bresciani2025amplitudes,bresciani2025positivity,Remmen:2020vts,Remmen:2020uze,Remmen:2022orj,duaso2025perturbative,cao2025unitarity,aoude2019rise}. We also apply spin-summing sum rules following Ref.~\cite{Remmen:2022orj} that exploit forward-scattering dispersion relations to determine whether the UV completion is dominated by scalar or vector intermediate states; this analysis is subject to standard caveats such as loop-level UV completions, non-trivial interplay between spin-0 and spin-1 exchange, and the assumption of vanishing boundary terms in the dispersion relation~\cite{Remmen:2022orj,Adams:2006sv,deRham:2017avq,Bellazzini:2014waa,Low:2009di,Distler:2006if,Chala:2021pll,Zhang:2020jyn}. The sign of the Wilson coefficient carries a memory of the dominant spin in the UV. These results allow us to partition the Wilson-coefficient parameter space into scalar-dominated and vector-dominated regions.

We then compare these theory bounds to the experimental constraints from NA64$\mu$ and the global fit of Ref.~\cite{Falkowski:2017pss}, along with the unitarity limit from perturbative unitarity~\cite{Lee:1977yc,Corbett:2017qgl}. Finally, we consider a concrete UV completion of these operators in the form 
of a leptophilic $Z'$ boson. The idea of gauging differences of lepton-family 
numbers was introduced in the early papers of He, Joshi, Lew, and 
Volkas~\cite{He:1991qd,Foot:1994vd}, who pointed out that $L_\mu - L_\tau$ 
is anomaly-free in the SM without adding any new fermions, making the 
corresponding $Z'$ one of the most minimal and well-motivated extensions of 
the SM. This model has since attracted renewed interest as a portal to dark 
matter, as an explanation of the muon anomalous magnetic moment, and as 
a source of neutrino self-interactions relevant for cosmological 
tensions~\cite{Escudero:2018mvt,Altmannshofer:2019zhy,Crivellin:2020ebi}. 
Recent global analyses and collider 
studies~\cite{buras2021global,suarez2025leptophilic} have mapped 
out the current experimental constraints on the $(g',\, M_{Z'})$ parameter 
space. Under the heavy-$Z'$ assumption, the SMEFT Wilson coefficients map 
directly onto this parameter space. We have also examined the JUNO projection 
in this parameter space for a first-generation benchmark, taken to be the 
kinetic-mixing dark photon model of Ref.~\cite{bauer2018hunting}.

The paper is organized as follows. In Sec.~\ref{sec:theory} we review perturbative unitarity bounds and spin-summing positivity constraints. In Sec.~\ref{sec:experiment} we present the experimental bounds, discuss RG running, and compare theory and experiment. In Sec.~\ref{sec:zprime} we compute $G_{\rm eff}$ from the Wilson coefficients and translate the bounds into constraints on a $Z'$ UV model. We conclude in Sec.~\ref{sec:conclusions}.

\section{Theoretical bounds}
\label{sec:theory}

\subsection{Perturbative unitarity}

For a $2\!\to\!2$ process the partial-wave expansion of the amplitude in the center-of-mass frame reads~\cite{Jacob:1959at,Itzykson:1980rh}
\begin{equation}
\mathcal{M}(\theta)=16\pi\sum_{J}(2J+1)\,a^J\,d^J_{\lambda\lambda'}(\theta),
\end{equation}
where $d^J$ are the Wigner functions. Probability conservation enforces the elastic unitarity bound $|\mathrm{Re}\,a^J|\le 1/2$~\cite{Lee:1977yc,Froissart:1961ux,Martin:1962rt}. For a contact dimension-six four-fermion interaction with coefficient $C/\Lambda^2$, the leading high-energy behavior of the $s$-channel amplitude is
\begin{equation}
\mathcal{M} \sim \frac{C}{\Lambda^2}\,s,
\end{equation}
and projecting onto the $J=0$ partial wave yields~\cite{bresciani2026unitarity}
\begin{equation}
\frac{|C|}{\Lambda^2}\lesssim \frac{2\pi}{s},
\label{eq:unitarity}
\end{equation}
for the marginalized $(\bar L L)(\bar L L)$ coefficient of $\mathcal{O}_{\ell\ell}$, where the marginalization corresponds to averaging over the helicity configurations that contribute to the $J=0$ singlet. The corresponding marginalized bound for the $(\bar L L)(\bar R R)$ coefficient of $\mathcal{O}_{\ell e}$ is $4\pi/s$, a factor of two weaker. We use this leading high-energy form throughout, neglecting the interference with SM amplitudes; a full helicity-resolved treatment would modify the numerical factors but not the qualitative comparison with experimental bounds. Equation~\eqref{eq:unitarity} defines a saturation scale $\sqrt{s_*}=\sqrt{2\pi\Lambda^2/|C|}$ above which the EFT description breaks down.

\subsection{Spin-summing positivity bounds}

\begin{figure}[t!]
\centering
\includegraphics[width=\columnwidth]{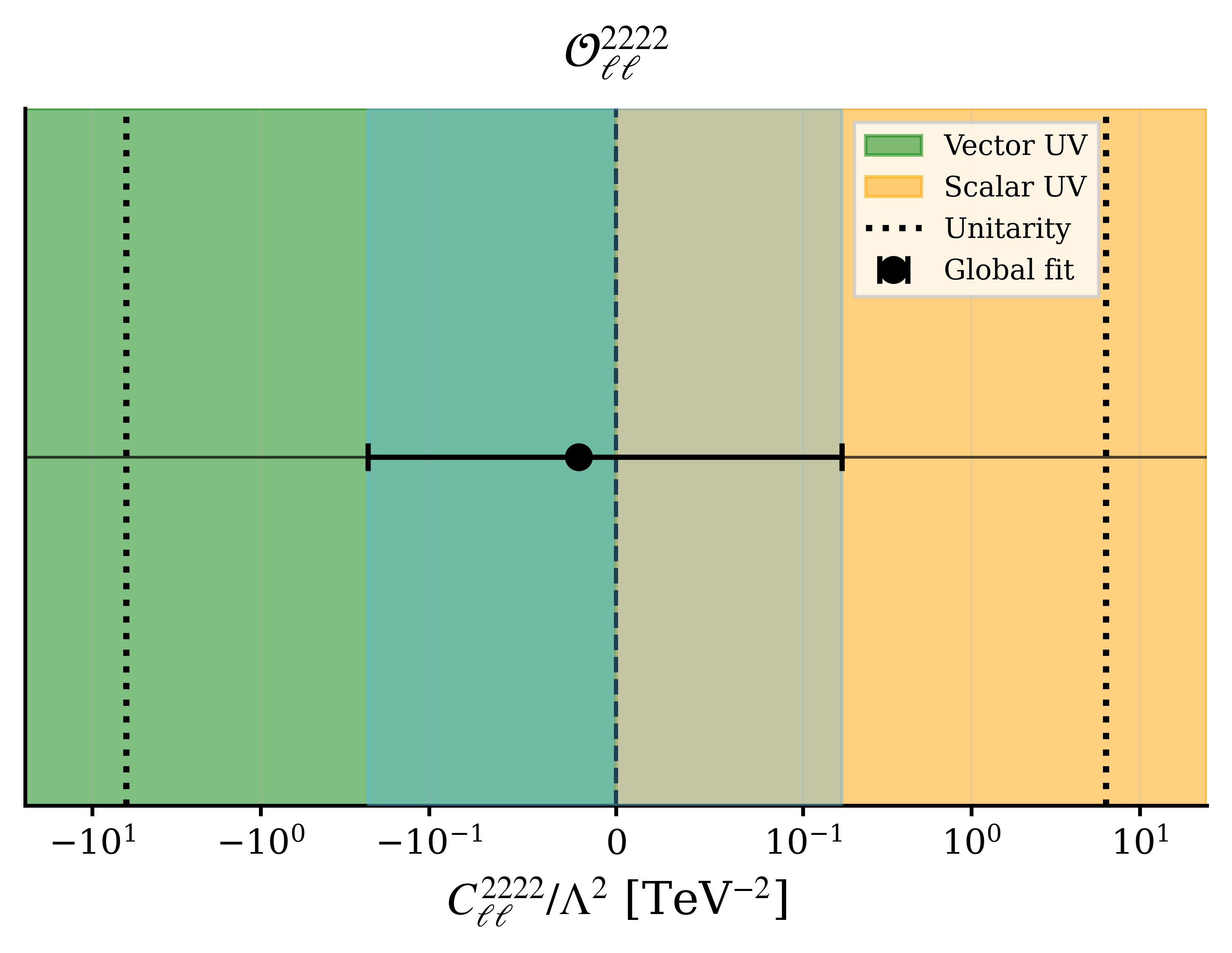}
\caption{Scalar- (orange) and vector-dominated (green) regions for $[C_{\ell\ell}]_{2222}$ from the spin-summing sum rules~\cite{Remmen:2022orj}, overlaid with the marginalized unitarity bound (dotted) and the global-fit central value~\cite{Falkowski:2017pss} under the assumption of a vanishing $C_{\ell e}$ contribution.}
\label{fig:spinsum}
\end{figure}

Forward-scattering dispersion relations combined with unitarity and crossing imply positivity constraints on the sign of dimension-six Wilson coefficients~\cite{Adams:2006sv,Low:2009di,Distler:2006if,Bellazzini:2014waa,deRham:2017avq,Remmen:2020vts,Chala:2021pll}. For SMEFT four-fermion operators, Remmen and Rodd showed~\cite{Remmen:2022orj} that
\begin{equation}
\lim_{s\to 0}\frac{\partial}{\partial s}\mathcal{A}_{\alpha\bar\beta}(s,0) = 8\int_{s_0}^{\infty}\frac{ds}{s^2}\Big[\mathrm{Im}\,a^{(0)}_{\alpha\bar\beta}(s)-3\,\mathrm{Im}\,a^{(1)}_{\alpha\beta}(s)\Big],
\label{eq:spinsum}
\end{equation}
so that a UV completion dominated by scalar intermediate states ($J=0$) makes the right-hand side positive, while a vector-dominated completion ($J=1$) makes it negative. For the operators considered here the relevant positive/negative combinations are $[C_{\ell\ell}]_{prst}$ and $[C_{\ell\ell}]_{prst}+[C_{\ell\ell}]_{ptsr}$, following Table~III of Ref.~\cite{Remmen:2022orj}. Concretely, scalar dominance implies $[C_{\ell\ell}]_{2222}>0$, $[C_{\ell\ell}]_{2233}>0$, and $[C_{\ell\ell}]_{2233}+[C_{\ell\ell}]_{2332}>0$, while vector dominance flips these signs. We will see this explicitly for the $L_\mu - L_\tau$ $Z'$ UV model in Sec.~\ref{sec:zprime}, where the tree-level matching produces diagonal coefficients consistent with vector dominance. Combining the spin-summing analysis with the global fit in Fig.~\ref{fig:spinsum} shows that the $[C_{\ell\ell}]_{2222}$ central value is compatible with either a scalar- or a vector-dominated UV completion within current uncertainties, so that positivity does not yet select a unique UV structure for this coefficient. For the $[C_{\ell\ell}]_{2222}$ analysis we assume the orthogonal $\mathcal{O}_{\ell e}$ contribution entering the hatted combination is zero, so that both the NA64$\mu$ and the global-fit constraints reduce to bounds on $[C_{\ell\ell}]_{2222}$ alone (see Sec.~\ref{sec:experiment}). We restrict here to the leading sign constraints derived from the forward-scattering dispersion relation; a full positivity polytope analysis~\cite{bresciani2025positivity} that exploits the matrix structure of the partial waves and yields correlated allowed regions is beyond the scope of the present work.

\section{Experimental bounds and comparison}
\label{sec:experiment}

\subsection{Global fit and NA64$\mu$}

Important constraints on four-lepton Wilson coefficients are provided by the global fit of Ref.~\cite{Falkowski:2017pss}, which incorporates data from LEP, $\tau$ decays, neutrino scattering, and parity-violation experiments. For the coefficients of interest, we quote
\begin{align}
\hat C_{\ell\ell}^{2222}/\Lambda^2 &= (-2\pm 21)\times 10^{-2}\;\text{TeV}^{-2},\\
[C_{\ell\ell}]_{2332}/\Lambda^2 &= (3\pm 2.3)\times 10^{-2}\;\text{TeV}^{-2},
\end{align}
while no direct bound exists for $[C_{\ell\ell}]_{2233}$ in this fit.

\begin{figure}[t!]
\centering
\includegraphics[width=\columnwidth]{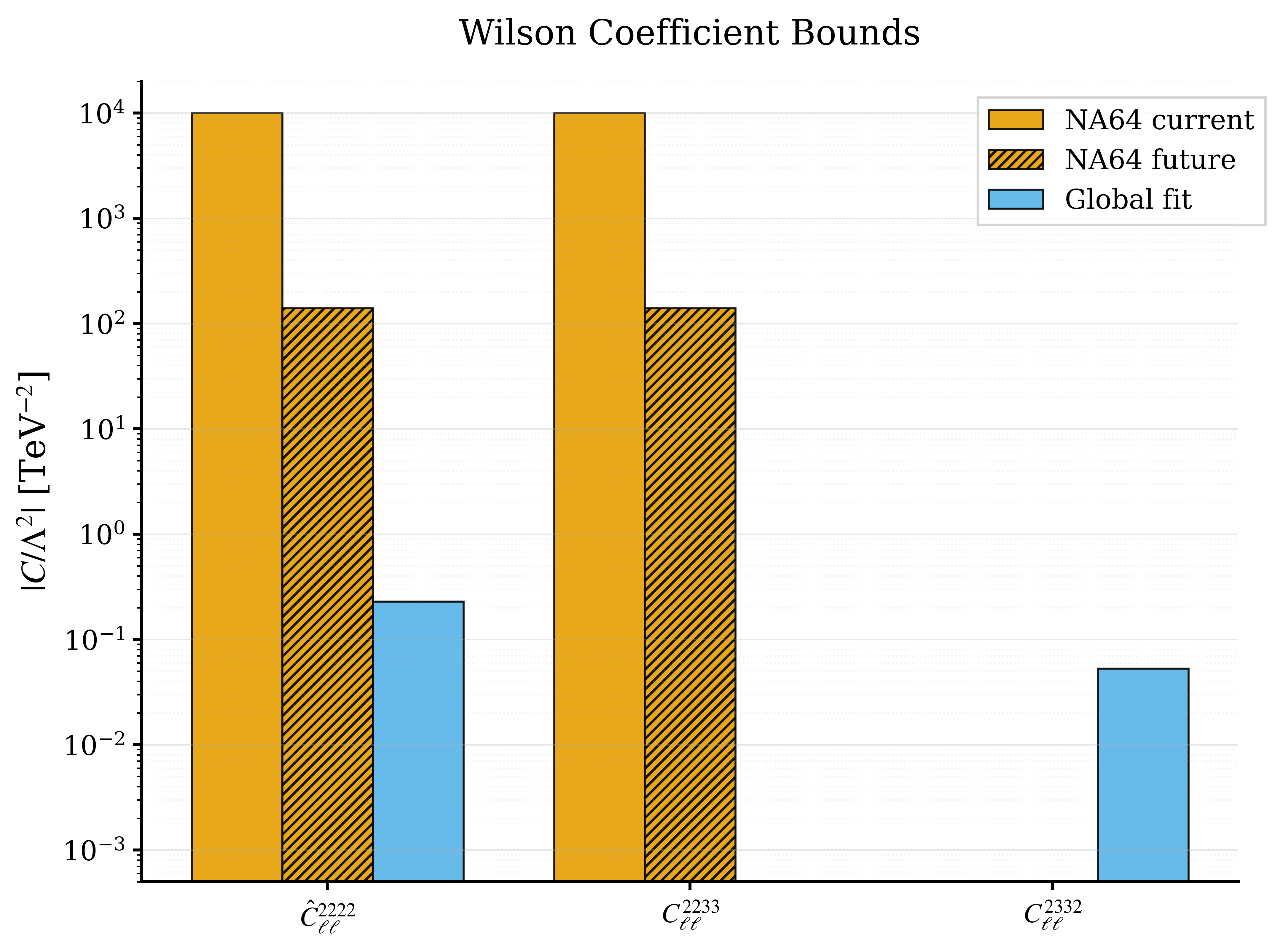}
\caption{Magnitude of the bounds on $\hat C_{\ell\ell}^{2222}$, $[C_{\ell\ell}]_{2233}$, and $[C_{\ell\ell}]_{2332}$ from NA64$\mu$ (current and projected)~\cite{NA64:2024nwj,crivelli2025first} and the global fit~\cite{Falkowski:2017pss}. Missing bars indicate no bound is available from the corresponding source.}
\label{fig:wilsoncomp}
\end{figure}

The NA64$\mu$ experiment at CERN~\cite{NA64:2024nwj,Gninenko:2014pea,crivelli2025first} produces muon beams on a fixed target and searches for missing-energy signatures. It is sensitive to the hatted combination $\hat C_{\ell\ell}^{2222}\equiv[C_{\ell\ell}]_{2222}+\frac{2 g_Y^2}{g_L^2+3g_Y^2}[C_{\ell e}]_{2222}$ via $\mu N\to\mu N\,\nu\bar\nu$ and to $[C_{\ell\ell}]_{2233}$ via $\mu N\to\mu N\,\nu_\tau\bar\nu_\tau$, where $g_L$ and $g_Y$ denote the SM $\mathrm{SU}(2)_L$ and $\mathrm{U}(1)_Y$ gauge couplings (with the numerical coefficient $\approx 0.32$, evaluated at the electroweak scale). The current and projected bounds translate to~\cite{crivelli2025first}
\begin{equation}
\frac{|\hat C_{\ell\ell}^{2222}|}{\Lambda^2},\;\frac{|[C_{\ell\ell}]_{2233}|}{\Lambda^2} \lesssim 
\begin{cases} 10^{4}\;\text{TeV}^{-2} & \text{(current)}, \\[2pt]
1.4\times 10^{2}\;\text{TeV}^{-2} & \text{(future)}.\end{cases}
\label{eq:NA64bound}
\end{equation}
Figure~\ref{fig:wilsoncomp} shows these current bounds and projected sensitivities for the three Wilson coefficients of interest. Identifying $\hat C_{\ell\ell}^{2222}\simeq[C_{\ell\ell}]_{2222}$ under the benchmark assumption $[C_{\ell e}]_{2222}=0$, the corresponding unitarity saturation scales from Eq.~\eqref{eq:unitarity} are $\sqrt{s_*}\simeq 25$~GeV (current) and $212$~GeV (projected), although direct collider searches probe a higher new-physics scale for these operators. We emphasize that this benchmark is a non-trivial assumption: the global fit of Ref.~\cite{Falkowski:2017pss} contains correlations between $[C_{\ell\ell}]_{2222}$ and $[C_{\ell e}]_{2222}$, and a non-zero $[C_{\ell e}]_{2222}$ would dilute the bound on $[C_{\ell\ell}]_{2222}$ along the NA64$\mu$ flat direction. The orthogonal direction $C_\perp$, defined explicitly below, is constrained only by perturbative unitarity. For $[C_{\ell\ell}]_{2233}$ this is the only experimental bound currently available, and it therefore sets the leading $\Lambda$ scale for that operator, as illustrated in Fig.~\ref{fig:energy}. Our saturation scales differ from the values quoted in Ref.~\cite{crivelli2025first} by an $\mathcal{O}(1)$ factor because we use the marginalized $(\bar L L)(\bar L L)$ bound rather than the individual partial-wave constraint; the two are equivalent statements about EFT validity but refer to different unitarity inequalities. In either case the saturation scales lie well above the typical NA64$\mu$ momentum transfer $\sqrt{q^2}\sim\mathcal{O}(\text{GeV})$, so $q^2/\Lambda^2 \ll 1$ throughout the fiducial region and the EFT description is self-consistent.

\subsection{RG running}

The Wilson coefficients evolve with the renormalization scale according to the SMEFT anomalous dimensions~\cite{Jenkins:2013zja,Jenkins:2013wua,Alonso:2013hga} and, below the electroweak scale, their LEFT counterparts~\cite{Jenkins:2017jig,Jenkins:2017dyc}. Purely leptonic $(\bar L L)(\bar L L)$ operators receive no QCD contribution, and the mixing into semi-leptonic and four-quark operators induced by electroweak loops is suppressed by additional gauge couplings and is numerically negligible for our purposes. Using the \texttt{wilson} package~\cite{Aebischer:2018bkb}, we evolve $[C_{\ell\ell}]_{2222,2233,2332}$ between $1$~GeV and $30$~TeV and find that running modifies the coefficients by $\lesssim 10\%$, consistent with prior expectations. We therefore neglect RG evolution as a subleading correction that does not change the interplay between unitarity, experiment, and UV sign constraints.

\subsection{Comparison}

\begin{figure}[t!]
\centering
\includegraphics[width=\columnwidth]{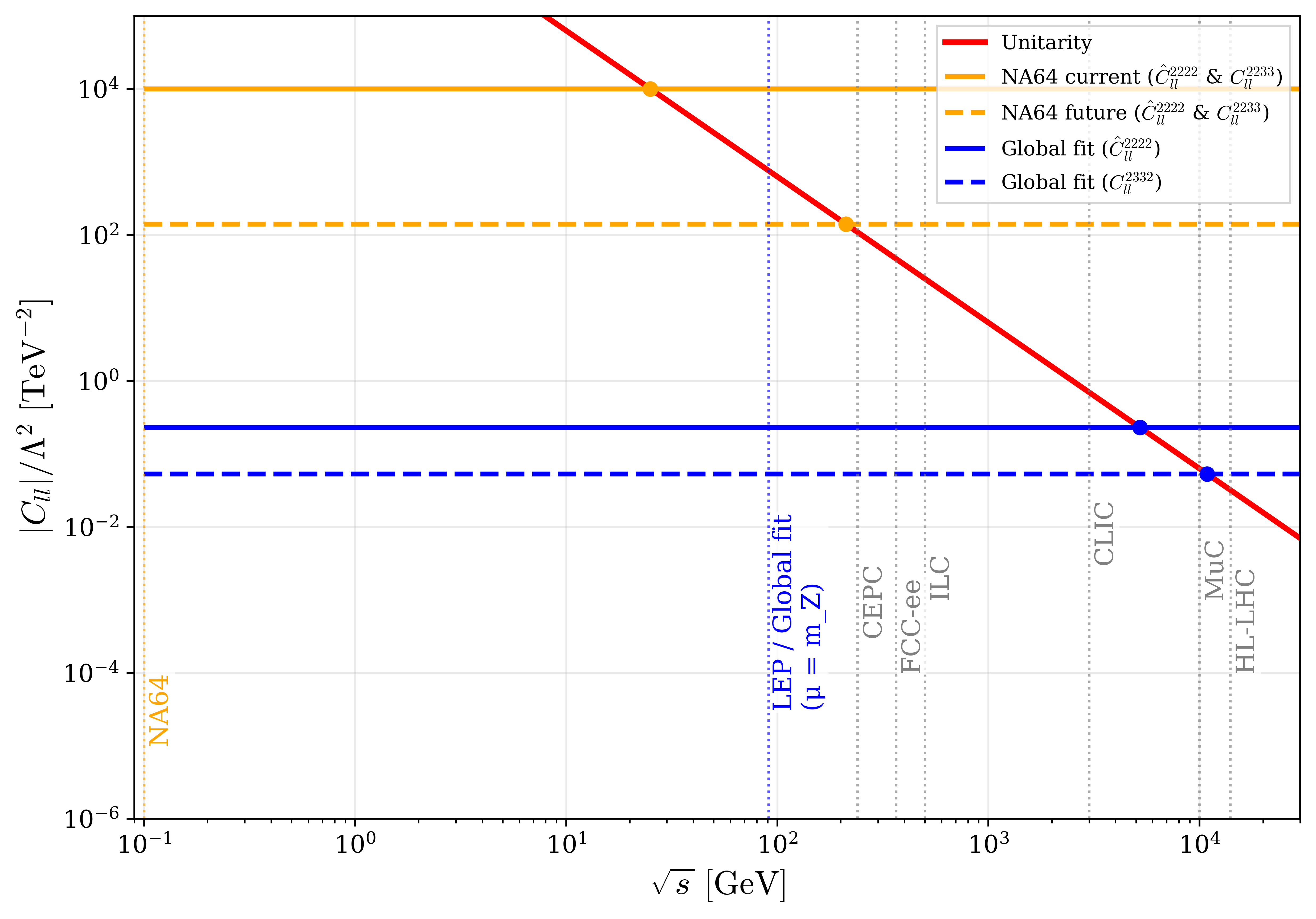}
\caption{Experimental bounds versus the perturbative-unitarity line $|C|/\Lambda^2=2\pi/s$ as a function of $\sqrt{s}$. The NA64$\mu$ future projection intersects the unitarity line at $\sqrt{s}\simeq 212$~GeV. Note the momentum transfer for the operator need not be the same as the experiment scale.}
\label{fig:energy}
\end{figure}

Figure~\ref{fig:energy} summarizes the interplay between experimental bounds and the unitarity line as a function of $\sqrt{s}$. The red line is Eq.~\eqref{eq:unitarity}, horizontal bands are the NA64$\mu$ current and projected bounds for $\hat C_{\ell\ell}^{2222}$ and $[C_{\ell\ell}]_{2233}$, and the blue lines are the global-fit limits on $\hat C_{\ell\ell}^{2222}$ and $[C_{\ell\ell}]_{2332}$. For reference we mark the center-of-mass energies of existing and proposed colliders such as LEP, FCC-ee~\cite{FCC:2018evy,deBlas:2019ehy}, CEPC~\cite{cepc2018cepc}, ILC~\cite{Bambade:2019fyw}, CLIC~\cite{CLICdp:2018cto}, a 10-TeV muon collider (MuC)~\cite{Accettura:2023ked,AlAli:2021let,MuonCollider:2022xlm}, and HL-LHC.

\begin{figure*}[t]
\centering
\begin{subfigure}{0.32\textwidth}
    \centering
    \includegraphics[width=\linewidth]{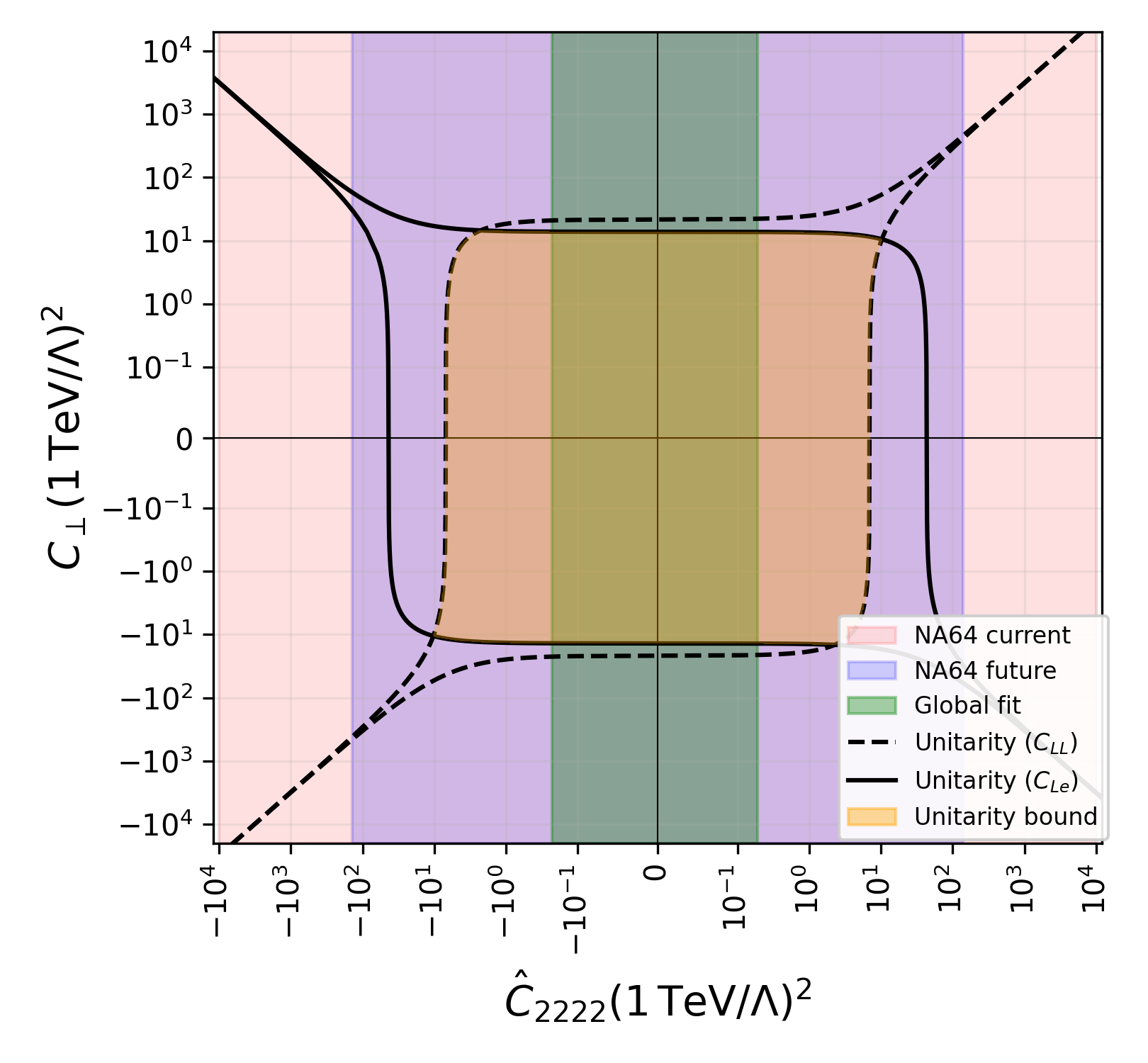}
    \caption{}
    \label{fig:chat-cperp}
\end{subfigure}
\hfill
\begin{subfigure}{0.32\textwidth}
    \centering
    \includegraphics[width=\linewidth]{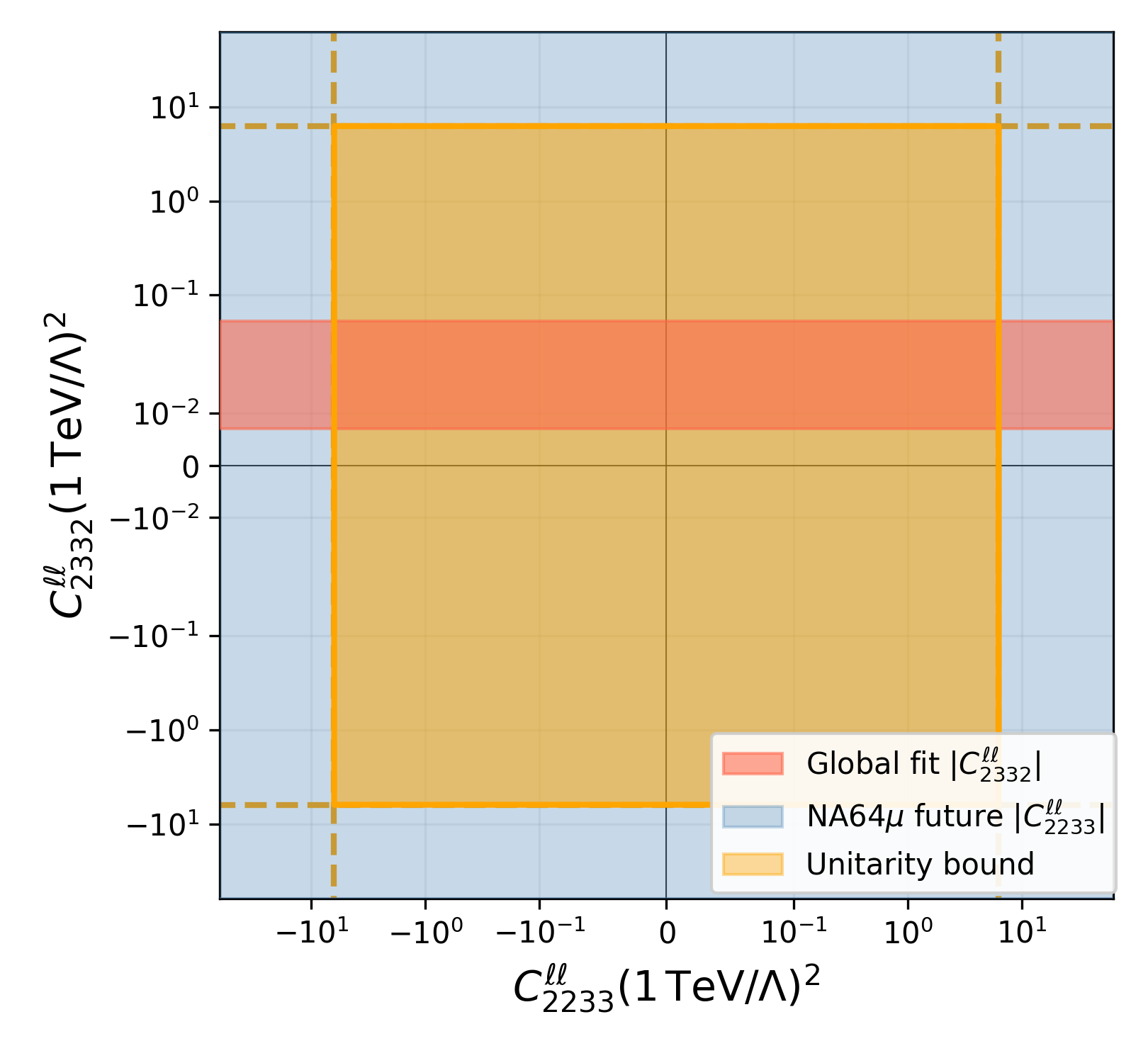}
    \caption{}
    \label{fig:c2233-c2332}
\end{subfigure}
\hfill
\begin{subfigure}{0.32\textwidth}
    \centering
    \includegraphics[width=\linewidth]{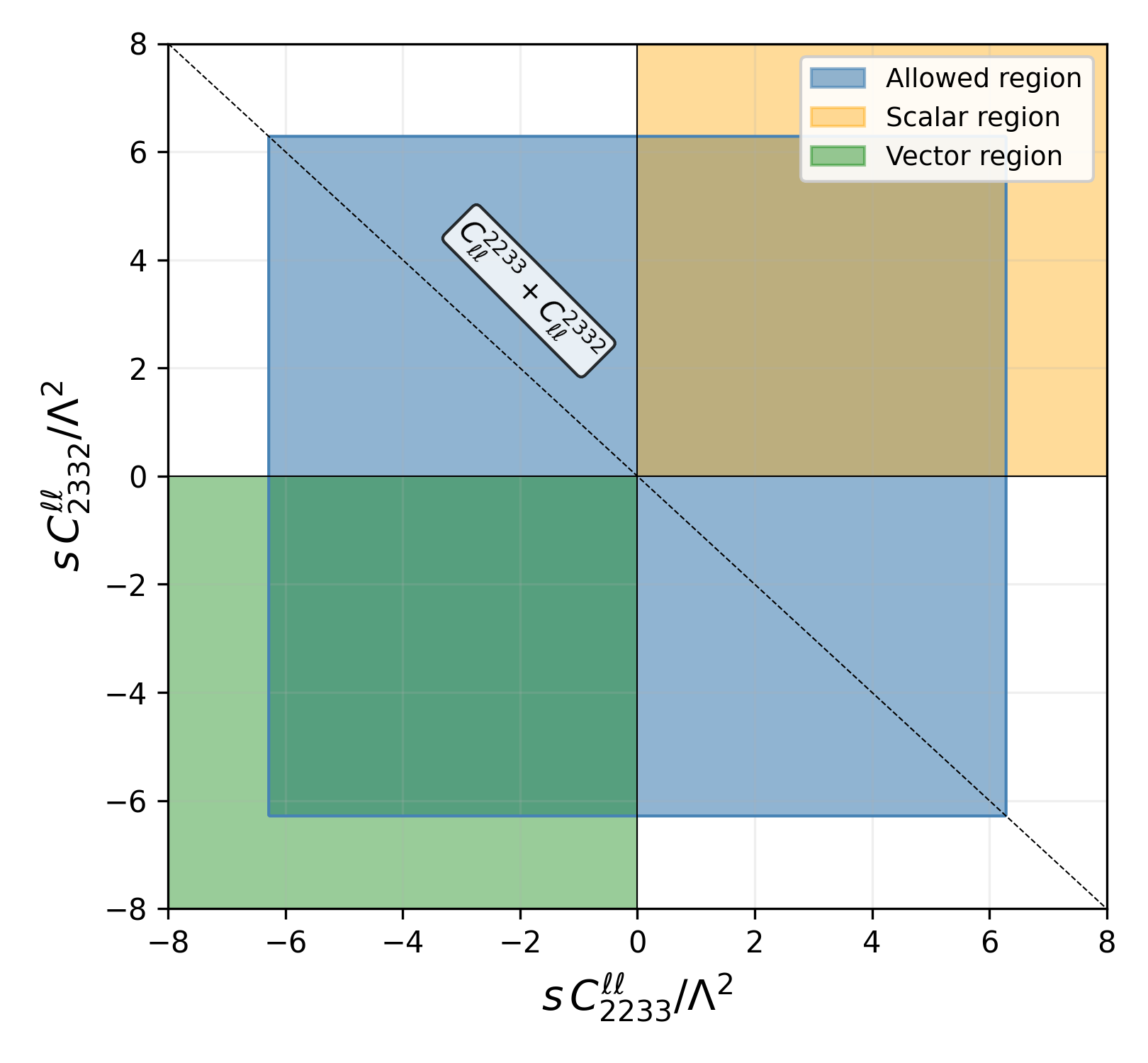}
    \caption{}
    \label{fig:spin-regions}
\end{subfigure}

\caption{
Combined experimental and theoretical constraints.
\textbf{(a)} Constraints in the rotated
\(\hat C_{\ell\ell}^{2222}\)--\(C_\perp\) plane. The dashed and solid
curves show the marginalized unitarity limits from
\(\mathcal O_{\ell\ell}\), \(2\pi\), and \(\mathcal O_{\ell e}\),
\(4\pi\), respectively.
\textbf{(b)} Allowed region in the
\([C_{\ell\ell}]_{2233}\)--\([C_{\ell\ell}]_{2332}\) plane, including
NA64\(\mu\) and global-fit constraints.
\textbf{(c)} Scalar- and vector UV regions inferred from
the spin-sum rule in Eq.~\eqref{eq:spinsum}. Here scalar (vector) regions are when both coefficients simultaneously are scalar (vector) UV completed.
}
\label{fig:final}
\end{figure*}

From the above analysis we draw the following conclusions:
\begin{itemize}
\item \emph{NA64$\mu$ provides the leading constraint on $[C_{\ell\ell}]_{2233}$}, since no other experiment currently bounds this coefficient directly.
\item \emph{The global fit dominates for $[C_{\ell\ell}]_{2332}$}, with present sensitivity few orders of magnitude below the unitarity line across all experimentally accessible $\sqrt{s}$.
\item \emph{For $[C_{\ell\ell}]_{2222}$ the global fit also dominates over NA64$\mu$}, modulo the flat-direction assumption $[C_{\ell e}]_{2222}=0$. The same conclusion holds for the hatted combination $\hat C_{\ell\ell}^{2222}$. Proposed $e^+e^-$ colliders (FCC-ee, ILC, CLIC) do not significantly improve the precision in the $\mu$--$\mu$ and $\mu$--$\tau$ sector.
\item Across the range of currently accessible energies (taken as $\sqrt{s}$ from the typical NA64$\mu$ momentum transfer up to the LEP scale and beyond), the perturbative unitarity bounds on $[C_{\ell\ell}]_{2222}$ and $[C_{\ell\ell}]_{2332}$ are weaker than the existing experimental constraints. In contrast, for $[C_{\ell\ell}]_{2233}$, which is currently weakly constrained, the unitarity bound enters the range of collider energies $\mathcal{O}(100)\,\mathrm{GeV}$. 
\end{itemize}
Future high-energy lepton colliders, such as a $\mu$ collider operating at $\sqrt{s} \sim 10~\mathrm{TeV}$~\cite{Accettura:2023ked,MuonCollider:2022xlm,AlAli:2021let}, can probe these operators at much higher energies. A high-energy muon collider would provide a direct probe of the
\(\mu\)--\(\tau\) four-lepton sector through
\(\mu^+\mu^-\to\tau^+\tau^-\), which receives contact-interaction
contributions from operators such as \([{\cal O}_{\ell\ell}]_{2233}\).
Since the corresponding dimension-six amplitude grows with energy as
\(\mathcal M_{\rm EFT}\sim (C/\Lambda^2)s\), higher center-of-mass
energies are expected to improve the sensitivity to these operators.
A quantitative projection, however, requires a dedicated collider
analysis including the Standard Model interference, angular cuts,
acceptances, luminosity, and systematic uncertainties, and is therefore
beyond the scope of this work. Future NA64$\mu$ improvements will also push the new-physics scale accessible to low-energy experiments to higher values. Together, these probes extend the reach of unitarity considerations and allow one to test the validity of the EFT description at significantly higher scales, potentially improving sensitivity to currently unconstrained or weakly constrained directions.

In Fig.~\ref{fig:final} we present the combined experimental and theoretical bounds in three coefficient planes. The left panel plots the unitarity and experimental bounds on the hatted combination $\hat{C}_{\ell\ell}^{2222}$ together with its orthogonal partner $C_\perp$. The direction $C_\perp$ is a flat direction for NA64$\mu$ but is bounded by perturbative unitarity. The unitarity bounds are tighter than NA64$\mu$ in this direction but less stringent than the global-fit bounds, as discussed above. Concretely, in the $(C_{\ell\ell}, C_{\ell e})$ plane we define the Euclidean-orthogonal coordinate:
\begin{equation}
C_\perp \equiv -\alpha\, [C_{\ell\ell}]_{2222} + [C_{\ell e}]_{2222}, \qquad \alpha \equiv \frac{2 g_Y^2}{g_L^2 + 3 g_Y^2} \approx 0.32,
\end{equation}
so that the basis $(\hat C, C_\perp)$ is orthogonal and the inverse transformation reads $[C_{\ell\ell}]_{2222} = (\hat C - \alpha C_\perp)/(1+\alpha^2)$ and $[C_{\ell e}]_{2222} = (\alpha \hat C + C_\perp)/(1+\alpha^2)$. The unitarity bounds $|[C_{\ell\ell}]_{2222}|/\Lambda^2 \le 2\pi/s$ and $|[C_{\ell e}]_{2222}|/\Lambda^2 \le 4\pi/s$ then translate to
\begin{align}
|\hat C - \alpha\, C_\perp| / \Lambda^2 &\le (1+\alpha^2)\, \frac{2\pi}{s}, \\
|\alpha\, \hat C + C_\perp| / \Lambda^2 &\le (1+\alpha^2)\, \frac{4\pi}{s},
\end{align}
which are the four lines bounding the allowed region in the left panel of Fig.~\ref{fig:final}. The center panel shows the perturbative-unitarity and experimental bounds in the $[C_{\ell\ell}]_{2233}$--$[C_{\ell\ell}]_{2332}$ plane. The right panel shows the unitarity region together with the scalar (orange) and vector (green) UV-dominated regions for the $\mathcal{O}_{\ell\ell}$ operator from Eq.~\eqref{eq:spinsum}, following the analysis of Ref.~\cite{bresciani2026unitarity} with scalar (vector) regions as the scenario when both the coefficients simultaneously are scalar (vector) UV completed.

\section{Implications for neutrino self-interactions and $Z'$ mediator}
\label{sec:zprime}

The SMEFT operators studied above generate $\nu$--$\nu$ scattering and therefore constrain the corresponding interaction strength. Upon matching the Warsaw-basis operator onto the low-energy LEFT, one obtains an effective four-neutrino interaction of the form
\begin{equation}
\mathcal{L}_{\nu\text{SI}} \supset G_{\rm eff}^{prst}\,(\bar\nu_p \gamma^\mu \nu_r)(\bar\nu_s \gamma_\mu \nu_t),
\end{equation}
which reduces to the schematic flavor-diagonal form $G_{\rm eff}^{ij}(\bar\nu_i\gamma^\mu\nu_i)(\bar\nu_j\gamma_\mu\nu_j)$ for $p=r=i$, $s=t=j$. For $E \ll m_Z$, the SM neutral current itself produces a four-neutrino contact interaction after integrating out the $Z$,
\begin{equation}
\mathcal{L}^{\nu\nu}_{\rm SM} = -\frac{G_F}{\sqrt{2}}\,\delta_{pr}\,\delta_{st}\,O^{prst}_{\nu\nu},
\end{equation}
with $O^{prst}_{\nu\nu}\equiv (\bar\nu_p\gamma^\mu\nu_r)(\bar\nu_s\gamma_\mu\nu_t)$. The general dimension-six four-neutrino operator generated by new physics is parametrized as
\begin{equation}
\mathcal{L}_{\rm EFT} = \frac{G_{\rm eff}^{prst}}{\sqrt{2}}\,O^{prst}_{\nu\nu}.
\end{equation}
Including both the SM contribution and the SMEFT matching at tree level, the Wilson coefficient of the $(V,LL)$ four-neutrino operator reads~\cite{Jenkins:2017jig,terol2020high}
\begin{multline}
C^{V,LL}_{\nu\nu}\big|_{prst} = \frac{[C_{\ell\ell}]_{prst}}{\Lambda^2} \\
- \frac{\bar g_Z^{\,2}}{4 m_Z^2}\Big([Z_\nu]_{pr}[Z_\nu]_{st}+[Z_\nu]_{pt}[Z_\nu]_{sr}\Big),
\label{eq:NSImatching}
\end{multline}
where $\bar g_Z$ and $[Z_\nu]_{ab}$ denote the effective $Z$-boson
coupling and the $Z\nu\nu$ coupling matrix after matching, including
possible SMEFT corrections. In the absence of new physics, they reduce
to $\bar g_Z=\sqrt{g_L^2+g_Y^2}$ and $[Z_\nu]_{ab}=\tfrac{1}{2}\,\delta_{ab}$.
Assuming no Higgs-sector correction, the Warsaw-basis operator $\mathcal{O}_{\ell\ell}$ generates an effective four-neutrino interaction whose strength is
\begin{equation}
G_{\rm eff}^{prst} = \sqrt{2}\,\frac{[C_{\ell\ell}]_{prst}}{\Lambda^2}.
\end{equation}

Using the global-fit and NA64$\mu$ bounds on the Wilson coefficients, this translates into the following limits on the effective neutrino self-coupling in the $\nu_\mu$--$\nu_\tau$ sector:
\begin{table}[h]
\centering
\scriptsize
\setlength{\tabcolsep}{4pt}
\begin{tabular}{lcc}
\toprule
Coefficient & Source & $|G_{\rm eff}|$~[GeV$^{-2}$] ($G_F$ ratio) \\
\midrule
$G_{\rm eff}^{2222}$ (global fit)         & \cite{Falkowski:2017pss} & $3.25\!\times\!10^{-7}$ ($\approx 2.8\!\times\!10^{-2}\, G_F$) \\
$G_{\rm eff}^{2233}$ (NA64$\mu$ current)  & \cite{crivelli2025first} & $1.41\!\times\!10^{-2}$ ($\approx 1.21\!\times\!10^{3}\, G_F$) \\
$G_{\rm eff}^{2233}$ (NA64$\mu$ future)   & \cite{crivelli2025first} & $1.97\!\times\!10^{-4}$ ($\approx 17\, G_F$) \\
$G_{\rm eff}^{2332}$ (global fit)         & \cite{Falkowski:2017pss} & $7.49\!\times\!10^{-8}$ ($\approx 6.46\!\times\!10^{-3}\, G_F$) \\
\bottomrule
\end{tabular}
\caption{Bounds on $G_{\rm eff}$ obtained by evaluating the Wilson coefficients at the upper edge of their quoted experimental ranges (central value plus uncertainty).}
\label{tab:Geff}
\end{table}

These laboratory bounds are many orders of magnitude smaller than the strongly interacting neutrino regime ($G_{\rm eff}\sim 10^{7}\text{--}10^{9}\,G_F$) proposed as a possible solution to the Hubble tension, and therefore exclude such large interaction strengths for the same operator and flavor structure. In particular, the mixed-flavor neutrino self-scattering channel $\nu_\mu\nu_\tau\to\nu_\mu\nu_\tau$ receives a direct SMEFT contribution from $[\mathcal{O}_{\ell\ell}]_{2233}$, which lacked a direct model-independent SMEFT bound until the NA64$\mu$ measurement~\cite{crivelli2025first}. The projected NA64$\mu$ sensitivity corresponds to a contact strength of order $\sim 17\,G_F$ (Table~\ref{tab:Geff}), which is many orders of magnitude smaller than the $G_{\rm eff}\sim 10^{7}\text{--}10^{9}\,G_F$ range invoked in strongly interacting neutrino cosmology, and therefore excludes such large interaction strengths. The corresponding first-generation interactions are constrained by the global analysis of Ref.~\cite{Falkowski:2017pss} to lie 2--3 orders of magnitude below $G_F$.

This hierarchy admits a clear physical interpretation. The strongly interacting neutrino regime, 
$G_{\rm eff} \sim 10^{7}$--$10^{9}\,G_F$, typically requires light mediators for which the interaction 
cannot be well approximated by a local contact operator. This regime, therefore, lies well outside 
the domain of validity of the dimension-six SMEFT framework.
By contrast, the bounds derived in this work apply to the contact-interaction regime, 
corresponding to heavy mediators with masses above the typical momentum transfer of the experiments. 
In this regime, the projected NA64$\mu$ sensitivity $G_{\rm eff}^{2233} \lesssim 17\,G_F$ 
probes mediator scales $\Lambda \gtrsim \mathcal{O}(100)\,\mathrm{MeV}$.
The global fit, which incorporates data from multiple experiments operating at different energies, 
yields significantly stronger constraints on $G_{\rm eff}^{2222}$ and $G_{\rm eff}^{2332}$.
These results exclude large contact-type $ \nu_{\mu}$-$\nu_{\mu}$ and $\nu_\mu$--$\nu_\tau$ self-interactions in the muon-tau sector 
within the validity of the dimension-six EFT. Light-mediator scenarios, such as those proposed 
to address the Hubble tension, remain viable precisely because they operate outside the contact-operator description. 
This motivates dedicated searches for light mediators at experiments such as NA64$\mu$, FASER, 
and other forward-physics facilities.
We note that higher-dimensional operators can become relevant when $s/\Lambda^2 \sim \mathcal{O}(1)$, 
and that many cosmological analyses invoke momentum-dependent or late-time interactions 
not captured by the simple contact-operator translation used here. Our conclusions should therefore 
be interpreted with these limitations in mind.

Under the assumption of a heavy-mediator UV completion (so that the SMEFT description is valid at the relevant momentum transfer), the NA64$\mu$ analysis on the Wilson coefficients can be translated into bounds on the parameter space of a chosen UV model, such as the muon $g-2$ motivated $L_\mu - L_\tau$ model. Since the SMEFT analysis is leading order, dedicated bounds from full NA64$\mu$ analyses for these models are also available in the literature; see Ref.~\cite{kelly2020origin} for a constraint on the $(g',\, m_{Z'})$ parameter space of the $L_\mu - L_\tau$ $Z'$ model from a range of experiments. Tree-level integration of a $Z'$ of mass $m_{Z'}$ and gauge coupling $g'$ generates the Wilson coefficients
\begin{align}
\frac{[C_{\ell\ell}]_{2222}}{\Lambda^2} &\sim -\frac{g'^2}{2 m_{Z'}^2}, & 
\frac{[C_{\ell\ell}]_{3333}}{\Lambda^2} &\sim -\frac{g'^2}{2 m_{Z'}^2}, \\
\frac{[C_{\ell\ell}]_{2233}}{\Lambda^2} &\sim +\frac{g'^2}{m_{Z'}^2}, & & \nonumber
\end{align}
with the $L_\mu - L_\tau$ current squaring out to give a cross-term coefficient with a sign opposite to the diagonal terms due to opposite charges. The $[\mathcal{O}_{\ell\ell}]_{2332}$ operator is not generated, since the flavor-conserving $L_\mu - L_\tau$ current involves no off-diagonal flavor structure. NA64$\mu$ provides direct sensitivity to $[\mathcal{O}_{\ell\ell}]_{2222}$ and $[\mathcal{O}_{\ell\ell}]_{2233}$, 
while the diagonal tau operator $[\mathcal{O}_{\ell\ell}]_{3333}$ remains essentially unconstrained 
without additional flavor assumptions. The negative sign for the diagonal coefficients realizes the vector-dominated regime expected from the spin-summing analysis of Sec.~\ref{sec:theory}. The bounds obtained in our SMEFT analysis, both for the current data and the projection, are similar to the dedicated analyses of Refs.~\cite{bauer2018hunting,kelly2020origin} for $m_{Z'}$ greater than the experimental momentum-transfer scale.

A similar analysis for the first-generation coefficients contributing to $\nu$--$\nu$ scattering, using projected JUNO sensitivities, is presented in Appendix~\ref{app:A}. JUNO's large fiducial volume enables stringent constraints on the Wilson coefficients of the $e$--$\mu$ sector~\cite{du2022exploring}. The nearby TAO detector further reduces the uncertainty in the unoscillated reactor antineutrino flux, improving the sensitivity to the relevant coefficients.

\section{Conclusions}
\label{sec:conclusions}

We compared three classes of bounds on the dimension-six four-lepton SMEFT 
operators in the $\mu$--$\tau$ sector. We derive bounds using perturbative unitarity from $2\to 2$ partial waves, spin-summing positivity sum rules, and direct experimental constraints from NA64$\mu$ together with a global fit to available data.

The sectoral hierarchy is robust and will not be closed by the next generation 
of $e^+e^-$ colliders. For $[C_{\ell\ell}]_{2222}$ and $[C_{\ell\ell}]_{2332}$, 
the global fit is a few orders of magnitude tighter than the unitarity 
line over the experimentally accessible energy range. For $[C_{\ell\ell}]_{2233}$, 
NA64$\mu$ is the only direct experimental probe; its projected sensitivity 
intersects the unitarity line at $\sqrt{s}$ around $200~\mathrm{GeV}$, within 
reach of proposed future colliders.

The spin-summing positivity sum rules partition the parameter space into 
scalar- and vector-dominated UV regions. The $L_\mu - L_\tau$ $Z'$ 
completion sits cleanly in the vector-dominated region, as expected from the 
negative sign of the tree-level matching for the diagonal coefficients. The 
current global-fit central value for $[C_{\ell\ell}]_{2222}$ is compatible 
with both signs, so positivity does not yet pick out a unique UV structure 
for that coefficient. While the global-fit central value for 
$[C_{\ell\ell}]_{2332}$ favors a scalar-dominated completion, this coefficient 
is not generated by a flavor-conserving $Z'$, so the implication does not 
constrain the leptophilic $Z'$ models considered here.

RG running between $1~\mathrm{GeV}$ and $30~\mathrm{TeV}$ modifies these 
coefficients by $\lesssim 10\%$. We have neglected it at leading order 
with the understanding that this is a subleading correction. We translated 
the Wilson-coefficient bounds onto the effective four-neutrino contact 
coupling $G_\mathrm{eff}$. The values we obtain are many orders of magnitude 
below the $G_\mathrm{eff} \sim 10^7$--$10^9\,G_F$ regime that has been 
invoked to ease cosmological tensions; within the validity of the dimension-six 
SMEFT and for contact-type interactions, our bounds therefore exclude 
heavy-mediator UV completions of strong $\nu_\mu$--$\nu_\tau$ and $\nu_\mu$--$\nu_\mu$ self-interactions, 
while leaving the cosmologically motivated light-mediator scenarios 
unconstrained by this analysis.

Finally, we used the experimental bounds to comment on the $(g',\, M_{Z'})$ 
parameter space of a leptophilic $L_\mu - L_\tau$ UV completion. The matching 
is valid for $M_{Z'}$ larger than the typical NA64$\mu$ momentum transfer. 
The analogous analysis for the $e$--$\mu$ sector, using the projected JUNO 
sensitivity to the first-generation Wilson coefficient $[C_{\ell\ell}]_{1122}$, 
is given in Appendix~\ref{app:A}. The large fiducial volume of JUNO probes 
a higher new-physics scale via this coefficient. The reach is stronger in the electron sector because global fits yield tighter constraints on first-generation coefficients, which are expected to improve with future experiments.

Beyond the explicit $Z'$ completion, these $\mu$--$\tau$ Wilson coefficients 
(and their analogues in other flavor sectors) also enter dark-matter scenarios 
in which neutrino self-interactions are generated at one loop through 
DM--neutrino couplings~\cite{arguelles2017imaging,mondol2025road}. In such 
models the four-neutrino contribution scales as $1/\Lambda^4$ and is further 
suppressed by a loop factor relative to the tree-level case studied here, but 
the same Wilson coefficients can in principle be used to constrain those 
models. The $\mu$--$\tau$ four-lepton sector is thus a natural meeting point 
of cosmology, EFT-based theoretical constraints, and direct laboratory 
experiments. We hope that the analysis presented here motivates further work 
in this direction.

\begin{acknowledgments}
We thank Tirtha Sankar Ray and Ranjini Mondol for stimulating talks where this question originated, and Tirtha Sankar Ray for further correspondence. G.~D.\ was supported in part by the INFN research initiative Exploring New Physics (ENP) and by an IISc Visiting Professorship (2026). S.~K.~V.\ thanks IISc and ANRF (Project ``Nature of New Physics'') for support. A.~S.\ thanks IISc and CSIR, Govt.\ of India, for SRF fellowship No.\ 09/0079(15487)/2022-EMR-I. The code used for plot/data generation in this work will be made publicly available on GitHub at
\href{https://github.com/AadarshSingh0}{\faGithub}
under a Creative Commons (CC) license.
\end{acknowledgments}

\appendix

\section{First-generation projection}
\label{app:A}
\begin{figure}[ht!]
    \centering
    \includegraphics[width=1.1\linewidth]{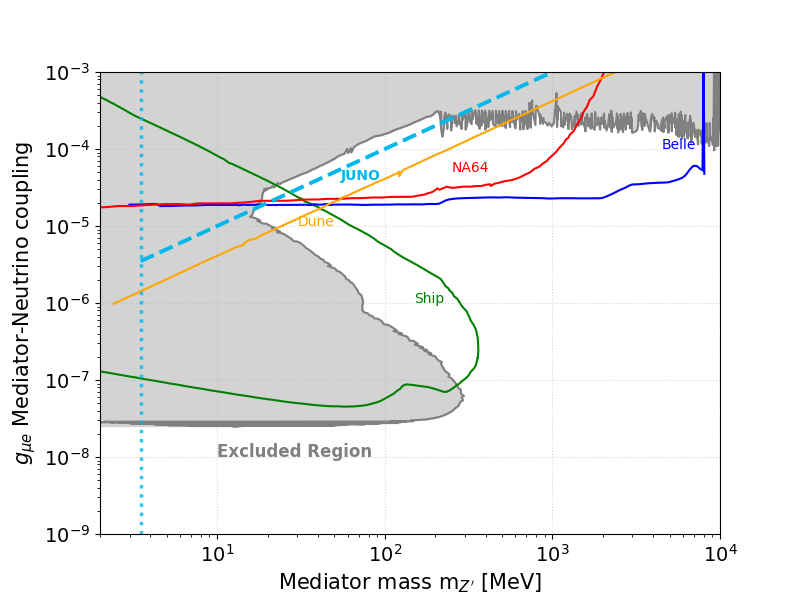}
    \caption{Constraints on the mediator--neutrino coupling versus mediator mass (in MeV). Existing bounds (white dwarf, BaBar, Borexino, and other probes), together with the Belle, NA64, DUNE, and SHiP projections, are adapted from Ref.~\cite{bauer2018hunting}; the JUNO projection (this work) is added using the projected sensitivity to the Wilson coefficient $[C_{\ell\ell}]_{1122}$.}
    \label{fig:e-mu}
\end{figure}
Beyond the $\mu$--$\tau$ analysis of Sec.~\ref{sec:zprime}, we can perform a similar analysis for the $e$--$\mu$ sector. Whereas the $L_\mu - L_\tau$ benchmark used in Sec.~\ref{sec:zprime}
gauges a flavor-non-universal current, the analogous $L_e - L_\mu$ symmetry is also anomaly-free with the
Standard Model fermion content, but is much
more strongly constrained by precision measurements involving electrons. We therefore adopt the kinetic-mixing dark-photon model of Ref.~\cite{bauer2018hunting} as our first-generation benchmark: a massive $Z'$ that mixes kinetically with the photon through the field-strength operator $-\tfrac{\epsilon}{2}\,F^{\mu\nu}F'_{\mu\nu}$. Since the electron sector is more strongly constrained at low energies and is also projected to receive the largest improvement from future experiments, the resulting parameter space is correspondingly more tightly constrained than in the $\mu$--$\tau$ case.

The matching procedure between the UV-model parameters and the Wilson coefficients is analogous to that of Sec.~\ref{sec:zprime}, with the kinetic-mixing parameter $\epsilon$ playing the role of the gauge coupling; we take $[C_{\ell\ell}]_{1122}$~\cite{du2022exploring} as the relevant coefficient for the projected bounds. JUNO is projected to reach a new-physics scale of $\sim 1$~TeV for this operator. The resulting projections are weaker than the corresponding projected DUNE bounds.

\bibliographystyle{unsrt}
\bibliography{bib}

@article{Grzadkowski:2010es,
  author = {Grzadkowski, B. and Iskrzynski, M. and Misiak, M. and Rosiek, J.},
  title = {Dimension-Six Terms in the {Standard Model Lagrangian}},
  journal = {JHEP},
  volume = {10},
  pages = {085},
  year = {2010},
  eprint = {1008.4884},
  archivePrefix = {arXiv},
  primaryClass = {hep-ph}
}

@article{Buchmuller:1985jz,
  author = {Buchmuller, W. and Wyler, D.},
  title = {Effective {Lagrangian} Analysis of New Interactions and Flavor Conservation},
  journal = {Nucl. Phys. B},
  volume = {268},
  pages = {621--653},
  year = {1986}
}

@article{Weinberg:1979sa,
  author = {Weinberg, Steven},
  title = {Phenomenological {Lagrangians}},
  journal = {Physica A},
  volume = {96},
  number = {1-2},
  pages = {327--340},
  year = {1979}
}

@article{Brivio:2017vri,
  author = {Brivio, Ilaria and Trott, Michael},
  title = {The Standard Model as an Effective Field Theory},
  journal = {Phys. Rept.},
  volume = {793},
  pages = {1--98},
  year = {2019},
  eprint = {1706.08945},
  archivePrefix = {arXiv},
  primaryClass = {hep-ph}
}

@article{Remmen:2020vts,
  author = {Remmen, Grant N. and Rodd, Nicholas L.},
  title = {Consistency of the {Standard Model} Effective Field Theory},
  journal = {JHEP},
  volume = {12},
  pages = {032},
  year = {2019},
  eprint = {1908.09845},
  archivePrefix = {arXiv},
  primaryClass = {hep-ph}
}

@article{Remmen:2020uze,
  author = {Remmen, Grant N. and Rodd, Nicholas L.},
  title = {Flavor Constraints from Unitarity and Analyticity},
  journal = {Phys. Rev. Lett.},
  volume = {125},
  number = {8},
  pages = {081601},
  year = {2020},
  eprint = {2004.02885},
  archivePrefix = {arXiv},
  primaryClass = {hep-ph},
  note = {[Erratum: Phys.Rev.Lett. 127, 149901 (2021)]}
}

@article{Remmen:2022orj,
  author = {Remmen, Grant N. and Rodd, Nicholas L.},
  title = {Spinning sum rules for the dimension-six {SMEFT}},
  journal = {JHEP},
  volume = {09},
  pages = {030},
  year = {2022},
  eprint = {2206.13524},
  archivePrefix = {arXiv},
  primaryClass = {hep-ph}
}

@article{Adams:2006sv,
  author = {Adams, Allan and Arkani-Hamed, Nima and Dubovsky, Sergei and Nicolis, Alberto and Rattazzi, Riccardo},
  title = {Causality, analyticity and an {IR} obstruction to {UV} completion},
  journal = {JHEP},
  volume = {10},
  pages = {014},
  year = {2006},
  eprint = {hep-th/0602178},
  archivePrefix = {arXiv}
}

@article{deRham:2017avq,
  author = {de Rham, Claudia and Melville, Scott and Tolley, Andrew J. and Zhou, Shuang-Yong},
  title = {Positivity bounds for scalar field theories},
  journal = {Phys. Rev. D},
  volume = {96},
  number = {8},
  pages = {081702},
  year = {2017},
  eprint = {1702.06134},
  archivePrefix = {arXiv},
  primaryClass = {hep-th}
}

@article{Falkowski:2017pss,
  author = {Falkowski, Adam and Gonzalez-Alonso, Martin and Mimouni, Kin},
  title = {Compilation of low-energy constraints on 4-fermion operators in the {SMEFT}},
  journal = {JHEP},
  volume = {08},
  pages = {123},
  year = {2017},
  eprint = {1706.03783},
  archivePrefix = {arXiv},
  primaryClass = {hep-ph}
}

@article{Jenkins:2013zja,
  author = {Jenkins, Elizabeth E. and Manohar, Aneesh V. and Trott, Michael},
  title = {Renormalization Group Evolution of the {Standard Model} Dimension Six Operators {I}: Formalism and $\lambda$ Dependence},
  journal = {JHEP},
  volume = {10},
  pages = {087},
  year = {2013},
  eprint = {1308.2627},
  archivePrefix = {arXiv},
  primaryClass = {hep-ph}
}

@article{Jenkins:2013wua,
  author = {Jenkins, Elizabeth E. and Manohar, Aneesh V. and Trott, Michael},
  title = {Renormalization Group Evolution of the {Standard Model} Dimension Six Operators {II}: {Yukawa} Dependence},
  journal = {JHEP},
  volume = {01},
  pages = {035},
  year = {2014},
  eprint = {1310.4838},
  archivePrefix = {arXiv},
  primaryClass = {hep-ph}
}

@article{Alonso:2013hga,
  author = {Alonso, Rodrigo and Jenkins, Elizabeth E. and Manohar, Aneesh V. and Trott, Michael},
  title = {Renormalization Group Evolution of the {Standard Model} Dimension Six Operators {III}: Gauge Coupling Dependence and Phenomenology},
  journal = {JHEP},
  volume = {04},
  pages = {159},
  year = {2014},
  eprint = {1312.2014},
  archivePrefix = {arXiv},
  primaryClass = {hep-ph}
}

@article{Aebischer:2018bkb,
  author = {Aebischer, Jason and Kumar, Jacky and Straub, David M.},
  title = {{Wilson}: a {Python} package for the running and matching of {Wilson} coefficients above and below the electroweak scale},
  journal = {Eur. Phys. J. C},
  volume = {78},
  number = {12},
  pages = {1026},
  year = {2018},
  eprint = {1804.05033},
  archivePrefix = {arXiv},
  primaryClass = {hep-ph}
}

@article{Jenkins:2017jig,
  author = {Jenkins, Elizabeth E. and Manohar, Aneesh V. and Stoffer, Peter},
  title = {Low-Energy Effective Field Theory below the Electroweak Scale: Operators and Matching},
  journal = {JHEP},
  volume = {03},
  pages = {016},
  year = {2018},
  eprint = {1709.04486},
  archivePrefix = {arXiv},
  primaryClass = {hep-ph}
}

@article{Jenkins:2017dyc,
  author = {Jenkins, Elizabeth E. and Manohar, Aneesh V. and Stoffer, Peter},
  title = {Low-Energy Effective Field Theory below the Electroweak Scale: Anomalous Dimensions},
  journal = {JHEP},
  volume = {01},
  pages = {084},
  year = {2018},
  eprint = {1711.05270},
  archivePrefix = {arXiv},
  primaryClass = {hep-ph}
}

@article{NA64:2024nwj,
  title={First results in the search for dark sectors at {NA64} with the {CERN SPS} high energy muon beam},
  author={Andreev, Yu M and Banerjee, Dipanwita and Banto Oberhauser, Benjamin and Bernhard, Johannes and Bisio, P and Charitonidis, Nikolaos and Crivelli, Paolo and Depero, Emilio and Dermenev, Alexander V and Donskov, Sergey V and others},
  journal={Physical Review Letters},
  volume={132},
  number={21},
  pages={211803},
  year={2024},
  publisher={APS}
}

@article{Gninenko:2014pea,
  author = {Gninenko, S. N. and Krasnikov, N. V. and Matveev, V. A.},
  title = {Muon $g-2$ and searches for a new leptophobic sub-{GeV} dark boson in a missing-energy experiment at {CERN}},
  journal = {Phys. Rev. D},
  volume = {91},
  pages = {095015},
  year = {2015},
  eprint = {1412.1400},
  archivePrefix = {arXiv},
  primaryClass = {hep-ph}
}

@article{crivelli2025first,
  title={First bounds on effective muon interactions using the {NA64}$\mu$ experiment at {CERN}},
  author={Crivelli, Paolo and Hernandez-Garcia, Josu and Lopez-Pavon, Jacobo and Lozano, Victor Martin and Bueno, Laura Molina},
  journal={arXiv preprint arXiv:2511.11801},
  year={2025}
}

@article{Accettura:2023ked,
  author = {Accettura, Carlotta and others},
  title = {Towards a muon collider},
  journal = {Eur. Phys. J. C},
  volume = {83},
  number = {9},
  pages = {864},
  year = {2023},
  eprint = {2303.08533},
  archivePrefix = {arXiv},
  primaryClass = {physics.acc-ph}
}

@article{AlAli:2021let,
  author = {Al Ali, Hind and others},
  title = {The muon {Smasher}'s guide},
  journal = {Rept. Prog. Phys.},
  volume = {85},
  number = {8},
  pages = {084201},
  year = {2022},
  eprint = {2103.14043},
  archivePrefix = {arXiv},
  primaryClass = {hep-ph}
}

@article{MuonCollider:2022xlm,
  author = {Black, K. M. and others},
  title = {Muon Collider Forum report},
  journal = {JINST},
  volume = {19},
  number = {02},
  pages = {T02015},
  year = {2024},
  eprint = {2209.01318},
  archivePrefix = {arXiv},
  primaryClass = {hep-ex}
}

@article{FCC:2018evy,
  author = {Abada, A. and others},
  collaboration = {FCC},
  title = {{FCC-ee}: The Lepton Collider: Future Circular Collider Conceptual Design Report Volume 2},
  journal = {Eur. Phys. J. ST},
  volume = {228},
  number = {2},
  pages = {261--623},
  year = {2019}
}

@article{cepc2018cepc,
  title={{CEPC} conceptual design report: Volume 2-physics \& detector},
  author={CEPC Study Group and others},
  journal={arXiv preprint arXiv:1811.10545},
  year={2018}
}

@article{Bambade:2019fyw,
  author = {Bambade, Philip and others},
  title = {The {International Linear Collider}: A Global Project},
  year = {2019},
  eprint = {1903.01629},
  archivePrefix = {arXiv},
  primaryClass = {hep-ex}
}

@article{CLICdp:2018cto,
  author = {Charles, T. K. and others},
  collaboration = {CLICdp, CLIC},
  title = {The {Compact Linear Collider} ({CLIC}) - 2018 Summary Report},
  journal = {CERN Yellow Rep. Monogr.},
  volume = {2},
  pages = {1--98},
  year = {2018},
  eprint = {1812.06018},
  archivePrefix = {arXiv},
  primaryClass = {physics.acc-ph}
}

@article{Ellis:2020unq,
  author = {Ellis, John and Madigan, Maeve and Mimasu, Ken and Sanz, Veronica and You, Tevong},
  title = {Top, {Higgs}, Diboson and Electroweak Fit to the {Standard Model} Effective Field Theory},
  journal = {JHEP},
  volume = {04},
  pages = {279},
  year = {2021},
  eprint = {2012.02779},
  archivePrefix = {arXiv},
  primaryClass = {hep-ph}
}

@article{Ethier:2021bye,
  author = {Ethier, Jacob J. and others},
  collaboration = {SMEFiT},
  title = {Combined {SMEFT} interpretation of {Higgs}, diboson, and top quark data from the {LHC}},
  journal = {JHEP},
  volume = {11},
  pages = {089},
  year = {2021},
  eprint = {2105.00006},
  archivePrefix = {arXiv},
  primaryClass = {hep-ph}
}

@article{deBlas:2022qbs,
  author = {de Blas, Jorge and Du, Yong and Grojean, Christophe and Gu, Jiayin and Miralles, V\'{\i}ctor and Peskin, Michael E. and Tian, Junping and Vos, Marcel and Vryonidou, Eleni},
  title = {Global {SMEFT} Fits at Future Colliders},
  year = {2022},
  eprint = {2206.08326},
  archivePrefix = {arXiv},
  primaryClass = {hep-ph}
}

@article{deBlas:2019ehy,
  author = {de Blas, J. and others},
  title = {{Higgs} Boson Studies at Future Particle Colliders},
  journal = {JHEP},
  volume = {01},
  pages = {139},
  year = {2020},
  eprint = {1905.03764},
  archivePrefix = {arXiv},
  primaryClass = {hep-ph}
}

@article{Berthier:2015oma,
  author = {Berthier, Laure and Trott, Michael},
  title = {Consistent constraints on the {Standard Model} Effective Field Theory},
  journal = {JHEP},
  volume = {02},
  pages = {069},
  year = {2016},
  eprint = {1508.05060},
  archivePrefix = {arXiv},
  primaryClass = {hep-ph}
}

@article{Corbett:2014ora,
  author = {Corbett, Tyler and Eboli, O. J. P. and Gonzalez-Fraile, J. and Gonzalez-Garcia, M. C.},
  title = {Constraining anomalous {Higgs} interactions},
  journal = {Phys. Rev. D},
  volume = {86},
  pages = {075013},
  year = {2012},
  eprint = {1207.1344},
  archivePrefix = {arXiv},
  primaryClass = {hep-ph}
}

@article{Corbett:2017qgl,
  author = {Corbett, Tyler and Eboli, O. J. P. and Gonzalez-Garcia, M. C.},
  title = {Unitarity Constrains on Dimension-Six Operators},
  journal = {Phys. Rev. D},
  volume = {91},
  number = {3},
  pages = {035014},
  year = {2015},
  eprint = {1411.5026},
  archivePrefix = {arXiv},
  primaryClass = {hep-ph}
}

@article{duaso2025perturbative,
  title={Perturbative unitarity bounds from momentum-space entanglement},
  author={Duaso Pueyo, Carlos and Goodhew, Harry and McCulloch, Ciaran and Pajer, Enrico},
  journal={Journal of High Energy Physics},
  volume={2025},
  number={8},
  pages={1--57},
  year={2025},
  publisher={Springer}
}

@article{Lee:1977yc,
  author = {Lee, Benjamin W. and Quigg, C. and Thacker, H. B.},
  title = {Weak Interactions at Very High-Energies: The Role of the {Higgs} Boson Mass},
  journal = {Phys. Rev. D},
  volume = {16},
  pages = {1519},
  year = {1977}
}

@article{Froissart:1961ux,
  author = {Froissart, Marcel},
  title = {Asymptotic behavior and subtractions in the {Mandelstam} representation},
  journal = {Phys. Rev.},
  volume = {123},
  pages = {1053--1057},
  year = {1961}
}

@article{Martin:1962rt,
  author = {Martin, Andre},
  title = {Unitarity and high-energy behavior of scattering amplitudes},
  journal = {Phys. Rev.},
  volume = {129},
  pages = {1432--1436},
  year = {1963}
}

@article{Jacob:1959at,
  author = {Jacob, M. and Wick, G. C.},
  title = {On the general theory of collisions for particles with spin},
  journal = {Annals Phys.},
  volume = {7},
  pages = {404--428},
  year = {1959}
}

@book{Itzykson:1980rh,
  author = {Itzykson, C. and Zuber, J. B.},
  title = {{Quantum Field Theory}},
  publisher = {McGraw-Hill},
  year = {1980}
}

@article{Escudero:2018mvt,
  author = {Escudero, Miguel and Hooper, Dan and Krnjaic, Gordan and Pierre, Mathias},
  title = {Cosmology with A Very Light $L_\mu - L_\tau$ Gauge Boson},
  journal = {JHEP},
  volume = {03},
  pages = {071},
  year = {2019},
  eprint = {1901.02010},
  archivePrefix = {arXiv},
  primaryClass = {hep-ph}
}

@article{Blinov:2019gcj,
  author = {Blinov, Nikita and Kelly, Kevin J. and Krnjaic, Gordan and McDermott, Samuel D.},
  title = {Constraining the Self-Interacting Neutrino Interpretation of the {Hubble} Tension},
  journal = {Phys. Rev. Lett.},
  volume = {123},
  number = {19},
  pages = {191102},
  year = {2019},
  eprint = {1905.02727},
  archivePrefix = {arXiv},
  primaryClass = {astro-ph.CO}
}

@article{Kreisch:2019yzn,
  author = {Kreisch, Christina D. and Cyr-Racine, Francis-Yan and Dor\'e, Olivier},
  title = {Neutrino puzzle: Anomalies, interactions, and cosmological tensions},
  journal = {Phys. Rev. D},
  volume = {101},
  number = {12},
  pages = {123505},
  year = {2020},
  eprint = {1902.00534},
  archivePrefix = {arXiv},
  primaryClass = {astro-ph.CO}
}

@article{Berryman:2022hds,
  author = {Berryman, Jeffrey M. and others},
  title = {Neutrino self-interactions: A white paper},
  journal = {Phys. Dark Univ.},
  volume = {42},
  pages = {101267},
  year = {2023},
  eprint = {2203.01955},
  archivePrefix = {arXiv},
  primaryClass = {hep-ph}
}

@article{lyu2021self,
  title={Self-interacting neutrinos: Solution to {Hubble} tension versus experimental constraints},
  author={Lyu, Kun-Feng and Stamou, Emmanuel and Wang, Lian-Tao},
  journal={Physical Review D},
  volume={103},
  number={1},
  pages={015004},
  year={2021},
  publisher={APS}
}

@article{Planck:2018vyg,
  author = {Aghanim, N. and others},
  collaboration = {Planck},
  title = {Planck 2018 results. {VI}. {Cosmological} parameters},
  journal = {Astron. Astrophys.},
  volume = {641},
  pages = {A6},
  year = {2020},
  eprint = {1807.06209},
  archivePrefix = {arXiv},
  primaryClass = {astro-ph.CO},
  note = {[Erratum: Astron.Astrophys. 652, C4 (2021)]}
}

@article{Riess:2021jrx,
  author = {Riess, Adam G. and others},
  title = {A Comprehensive Measurement of the Local Value of the {Hubble} Constant with 1 km/s/{Mpc} Uncertainty from the {Hubble Space Telescope} and the {SH0ES} Team},
  journal = {Astrophys. J. Lett.},
  volume = {934},
  number = {1},
  pages = {L7},
  year = {2022},
  eprint = {2112.04510},
  archivePrefix = {arXiv},
  primaryClass = {astro-ph.CO}
}

@article{Altmannshofer:2019zhy,
  author = {Altmannshofer, Wolfgang and Gori, Stefania and Pospelov, Maxim and Yavin, Itay},
  title = {Neutrino Trident Production: A Powerful Probe of New Physics with Neutrino Beams},
  journal = {Phys. Rev. Lett.},
  volume = {113},
  pages = {091801},
  year = {2014},
  eprint = {1406.2332},
  archivePrefix = {arXiv},
  primaryClass = {hep-ph}
}

@article{Altmannshofer:2014pba,
  author = {Altmannshofer, W. and Gori, S. and Pospelov, M. and Yavin, I.},
  title = {Quark flavor transitions in $L_\mu - L_\tau$ models},
  journal = {Phys. Rev. D},
  volume = {89},
  pages = {095033},
  year = {2014},
  eprint = {1403.1269},
  archivePrefix = {arXiv},
  primaryClass = {hep-ph}
}

@article{CCFR:1991lpl,
  author = {Mishra, S. R. and others},
  collaboration = {CCFR},
  title = {Neutrino tridents and {$WZ$} interference},
  journal = {Phys. Rev. Lett.},
  volume = {66},
  pages = {3117--3120},
  year = {1991}
}

@article{Falkowski:2019hvp,
  author = {Falkowski, Adam and Gonz\'alez-Alonso, Mart\'\i{}n and Tabrizi, Zahra},
  title = {Reactor neutrino oscillations as constraints on Effective Field Theory},
  journal = {JHEP},
  volume = {05},
  pages = {173},
  year = {2019},
  eprint = {1901.04553},
  archivePrefix = {arXiv},
  primaryClass = {hep-ph}
}

@article{ParticleDataGroup:2022pth,
  author = {Workman, R. L. and others},
  collaboration = {Particle Data Group},
  title = {Review of Particle Physics},
  journal = {PTEP},
  volume = {2022},
  pages = {083C01},
  year = {2022}
}

@article{Han:2020dwo,
  title={Quark and gluon contents of a lepton at high energies},
  author={Han, Tao and Ma, Yang and Xie, Keping},
  journal={Journal of High Energy Physics},
  volume={2022},
  number={2},
  pages={154},
  year={2022},
  publisher={Springer}
}

@article{Buttazzo:2020uzc,
  author = {Buttazzo, Dario and Paradisi, Paride},
  title = {Probing the muon $g-2$ anomaly with the {Higgs} boson at a muon collider},
  journal = {Phys. Rev. D},
  volume = {104},
  number = {7},
  pages = {075021},
  year = {2021},
  eprint = {2012.02769},
  archivePrefix = {arXiv},
  primaryClass = {hep-ph}
}

@article{Chala:2021pll,
  title={Positivity bounds in the {Standard Model} effective field theory beyond tree level},
  author={Chala, Mikael and Santiago, Jose},
  journal={Physical Review D},
  volume={105},
  number={11},
  pages={L111901},
  year={2022},
  publisher={APS}
}

@article{Zhang:2020jyn,
  author = {Zhang, Cen and Zhou, Shuang-Yong},
  title = {Positivity bounds on vector boson scattering at the {LHC}},
  journal = {Phys. Rev. D},
  volume = {100},
  number = {9},
  pages = {095003},
  year = {2019},
  eprint = {1808.00010},
  archivePrefix = {arXiv},
  primaryClass = {hep-ph}
}

@article{Bellazzini:2014waa,
  author = {Bellazzini, Brando and Martucci, Luca and Torre, Riccardo},
  title = {Symmetries, Sum Rules and Constraints on Effective Field Theories},
  journal = {JHEP},
  volume = {09},
  pages = {100},
  year = {2014},
  eprint = {1405.2960},
  archivePrefix = {arXiv},
  primaryClass = {hep-th}
}

@article{Low:2009di,
  author = {Low, Ian and Rattazzi, Riccardo and Vichi, Alessandro},
  title = {Theoretical Constraints on the {Higgs} Effective Couplings},
  journal = {JHEP},
  volume = {04},
  pages = {126},
  year = {2010},
  eprint = {0907.5413},
  archivePrefix = {arXiv},
  primaryClass = {hep-ph}
}

@article{Distler:2006if,
  author = {Distler, Jacques and Grinstein, Benjamin and Porto, Rafael A. and Rothstein, Ira Z.},
  title = {Falsifying Models of New Physics via {$WW$} Scattering},
  journal = {Phys. Rev. Lett.},
  volume = {98},
  pages = {041601},
  year = {2007},
  eprint = {hep-ph/0604261},
  archivePrefix = {arXiv}
}

@article{cao2025unitarity,
  title={Unitarity bounds and basis transformations in {SMEFT}: An analysis of {Warsaw} and {SILH} bases},
  author={Cao, Qing-Hong and Liu, Yandong and Yuan, Shu-Run},
  journal={Nuclear Physics B},
  volume={1010},
  pages={116781},
  year={2025},
  publisher={Elsevier}
}

@article{LEP:2003aa,
  author = {Schael, S. and others},
  collaboration = {ALEPH, DELPHI, L3, OPAL, LEP Electroweak},
  title = {Electroweak Measurements in Electron-Positron Collisions at {W}-Boson-Pair Energies at {LEP}},
  journal = {Phys. Rept.},
  volume = {532},
  pages = {119--244},
  year = {2013},
  eprint = {1302.3415},
  archivePrefix = {arXiv},
  primaryClass = {hep-ex}
}

@article{Efrati:2015eaa,
  author = {Efrati, Aielet and Falkowski, Adam and Soreq, Yotam},
  title = {Electroweak constraints on flavorful effective theories},
  journal = {JHEP},
  volume = {07},
  pages = {018},
  year = {2015},
  eprint = {1503.07872},
  archivePrefix = {arXiv},
  primaryClass = {hep-ph}
}

@article{Crivellin:2020ebi,
  author = {Crivellin, Andreas and Mellado, Bruce},
  title = {Anomalies in particle physics and their implications for physics beyond the {Standard Model}},
  journal = {Nature Rev. Phys.},
  volume = {6},
  number = {5},
  pages = {294--309},
  year = {2024},
  eprint = {2309.03870},
  archivePrefix = {arXiv},
  primaryClass = {hep-ph}
}

@article{Dedes:2017zog,
  author = {Dedes, Athanasios and Materkowska, W. and Paraskevas, M. and Rosiek, J. and Suxho, K.},
  title = {Feynman Rules for the {Standard Model} Effective Field Theory in {$R_\xi$-gauges}},
  journal = {JHEP},
  volume = {06},
  pages = {143},
  year = {2017},
  eprint = {1704.03888},
  archivePrefix = {arXiv},
  primaryClass = {hep-ph}
}

@article{aoude2019rise,
  title={The rise of {SMEFT} on-shell amplitudes},
  author={Aoude, Rafael and Machado, Camila S},
  journal={Journal of High Energy Physics},
  volume={2019},
  number={12},
  pages={58},
  year={2019},
  publisher={Springer}
}

@article{brinckmann2021self,
  title={Self-interacting neutrinos, the {Hubble} parameter tension, and the cosmic microwave background},
  author={Brinckmann, Thejs and Chang, Jae Hyeok and LoVerde, Marilena},
  journal={Physical Review D},
  volume={104},
  number={6},
  pages={063523},
  year={2021},
  publisher={APS}
}

@article{mondol2025road,
  title={Road through Dark$\nu$ess: Probing dark matter--neutrino interactions using {KM3-230213A}},
  author={Mondol, Ranjini and Bouri, Subhadip and Saha, Akash Kumar and Laha, Ranjan},
  journal={arXiv preprint arXiv:2506.19910},
  year={2025}
}

@article{arguelles2017imaging,
  title={Imaging galactic dark matter with high-energy cosmic neutrinos},
  author={Arg{\"u}elles, Carlos A and Kheirandish, Ali and Vincent, Aaron C},
  journal={Physical Review Letters},
  volume={119},
  number={20},
  pages={201801},
  year={2017},
  publisher={APS}
}

@article{adam2015juno,
  title={{JUNO} conceptual design report},
  author={Adam, T and An, F and An, G and An, Q and Anfimov, N and Antonelli, V and Baccolo, G and Baldoncini, M and Baussan, E and Bellato, M and others},
  journal={arXiv preprint arXiv:1508.07166},
  year={2015}
}

@article{an2016juno,
  title={Neutrino physics with {JUNO}},
  author={An, Fengpeng and others},
  collaboration={JUNO},
  journal={Journal of Physics G: Nuclear and Particle Physics},
  volume={43},
  number={3},
  pages={030401},
  year={2016},
  publisher={IOP Publishing},
  eprint={1507.05010},
  archivePrefix={arXiv}
}

@article{km3net2021orca,
  title={Determining the neutrino mass ordering and oscillation parameters 
         with {KM3NeT/ORCA}},
  author={Aiello, S and others},
  collaboration={KM3NeT},
  journal={European Physical Journal C},
  volume={81},
  pages={1008},
  year={2021},
  eprint={2103.09885},
  archivePrefix={arXiv}
}

@article{abusleme2020tao,
  title={{TAO} conceptual design report: A precision measurement of the reactor antineutrino spectrum with sub-percent energy resolution},
  author={Abusleme, Angel and Adam, Thomas and Ahmad, Shakeel and Aiello, Sebastiano and Akram, Muhammad and Ali, Nawab and An, Fengpeng and An, Guangpeng and An, Qi and Andronico, Giuseppe and others},
  journal={arXiv preprint arXiv:2005.08745},
  year={2020}
}

@article{abi2020deep,
  title={{Deep Underground Neutrino Experiment (DUNE)}, far detector technical design report, volume {II}: {DUNE} physics},
  author={Abi, Babak and Acciarri, Roberto and Acero, Mario A and Adamov, Giorge and Adams, David and Adinolfi, Marco and Ahmad, Zubayer and Ahmed, Jhanzeb and Alion, Tyler and Monsalve, S Alonso and others},
  journal={arXiv preprint arXiv:2002.03005},
  year={2020}
}

@article{abe2018hyper,
  title={{Hyper-Kamiokande} design report},
  author={Abe, Ke and Abe, Ke and Aihara, H and Aimi, A and Akutsu, R and Andreopoulos, C and Anghel, I and Anthony, LHV and Antonova, M and Ashida, Y and others},
  journal={arXiv preprint arXiv:1805.04163},
  year={2018}
}

@article{aartsen2018measurement,
  title={Measurement of atmospheric neutrino oscillations at 6--56 {GeV} with {IceCube DeepCore}},
  author={Aartsen, MG and Ackermann, M and Adams, J and Aguilar, JA and Ahlers, M and Ahrens, M and Al Samarai, I and Altmann, D and Andeen, K and Anderson, T and others},
  journal={Physical review letters},
  volume={120},
  number={7},
  pages={071801},
  year={2018},
  publisher={APS}
}

@article{adrian2016letter,
  title={Letter of intent for {KM3NeT} 2.0},
  author={Adrian-Martinez, Silvia and Ageron, M and Aharonian, F and Aiello, S and Albert, A and Ameli, F and Anassontzis, E and Andre, M and Androulakis, G and Anghinolfi, M and others},
  journal={Journal of Physics G: Nuclear and Particle Physics},
  volume={43},
  number={8},
  pages={084001},
  year={2016},
  publisher={IoP Publishing}
}

@article{du2022exploring,
  title={Exploring {SMEFT}-induced nonstandard interactions: From {COHERENT} to neutrino oscillations},
  author={Du, Yong and Li, Hao-Lin and Tang, Jian and Vihonen, Sampsa and Yu, Jiang-Hao},
  journal={Physical Review D},
  volume={105},
  number={7},
  pages={075022},
  year={2022},
  publisher={APS}
}

@article{das2022selfinteracting,
  author = {Das, Anirban},
  title = {Self-interacting neutrinos as a solution to the {Hubble} tension?},
  year = {2022},
  eprint = {2109.03263},
  archivePrefix = {arXiv},
  primaryClass = {hep-ph}
}

@article{roychoudhury2021updated,
  author = {Roy Choudhury, Shouvik and Hannestad, Steen and Tram, Thomas},
  title = {Updated constraints on massive neutrino self-interactions from cosmology in light of the {$H_0$} tension},
  journal = {JCAP},
  volume = {03},
  pages = {084},
  year = {2021},
  eprint = {2012.07519},
  archivePrefix = {arXiv},
  primaryClass = {astro-ph.CO}
}

@article{bresciani2026unitarity,
  title={Unitarity bounds and sum rules in the {SMEFT}},
  author={Bresciani, Luigi C and Paradisi, Paride and Sainaghi, Andrea},
  journal={arXiv preprint arXiv:2603.03423},
  year={2026}
}

@article{bresciani2025amplitudes,
  title={Amplitudes and partial wave unitarity bounds},
  author={Bresciani, Luigi C and Levati, Gabriele and Paradisi, Paride},
  journal={arXiv preprint arXiv:2504.12855},
  year={2025}
}

@article{bresciani2025positivity,
  title={Positivity and partial wave unitarity bounds on {ALP} theories via amplitude methods},
  author={Bresciani, Luigi C and Levati, Gabriele and Paradisi, Paride},
  journal={arXiv preprint arXiv:2510.13953},
  year={2025}
}

@article{He:1991qd,
  author = {He, X. G. and Joshi, G. C. and Lew, H. and Volkas, R. R.},
  title = {Simplest {$Z'$} model},
  journal = {Phys. Rev. D},
  volume = {44},
  pages = {2118--2132},
  year = {1991}
}

@article{Foot:1994vd,
  author = {Foot, R. and He, X. G. and Lew, H. and Volkas, R. R.},
  title = {Model for a light {$Z'$} boson},
  journal = {Phys. Rev. D},
  volume = {50},
  pages = {4571--4580},
  year = {1994},
  eprint = {hep-ph/9401250},
  archivePrefix = {arXiv}
}

@article{buras2021global,
  title={Global analysis of leptophilic {$Z'$} bosons},
  author={Buras, Andrzej J and Crivellin, Andreas and Kirk, Fiona and Manzari, Claudio Andrea and Montull, Marc},
  journal={Journal of High Energy Physics},
  volume={2021},
  number={6},
  pages={1--44},
  year={2021},
  publisher={Springer}
}

@article{suarez2025leptophilic,
  title={Leptophilic {$Z'$} bosons at the {FCC-ee}: Discovery opportunities},
  author={Suarez, Rebeca Gonzalez and Pattnaik, Baibhab and Zurita, Jos{\'e}},
  journal={Physical Review D},
  volume={111},
  number={3},
  pages={035029},
  year={2025},
  publisher={APS}
}

@article{kelly2020origin,
  title={Origin of sterile neutrino dark matter via secret neutrino interactions with vector bosons},
  author={Kelly, Kevin J and Sen, Manibrata and Tangarife, Walter and Zhang, Yue},
  journal={Physical Review D},
  volume={101},
  number={11},
  pages={115031},
  year={2020},
  publisher={APS}
}

@article{bauer2018hunting,
  title={Hunting all the hidden photons},
  author={Bauer, Martin and Foldenauer, Patrick and Jaeckel, Joerg},
  journal={Journal of High Energy Physics},
  volume={2018},
  number={7},
  pages={1--47},
  year={2018},
  publisher={Springer}
}

@article{terol2020high,
  title={High-energy constraints from low-energy neutrino nonstandard interactions},
  author={Terol-Calvo, Jorge and T{\'o}rtola, Mariam and Vicente, Avelino},
  journal={Physical Review D},
  volume={101},
  number={9},
  pages={095010},
  year={2020},
  publisher={APS}
}

\end{document}